\documentclass[%
 reprint,
 amsmath,amssymb,
 aps,
 pra,
]{revtex4-1}

\usepackage{IEEEtrantools}
\usepackage{graphicx}
\usepackage{xspace}
\usepackage{soul}
\usepackage{siunitx}
\usepackage{tabularx}
\usepackage[nomargin,inline,draft]{fixme}
\fxusetheme{color}
\usepackage{dcolumn}% Align table columns on decimal point
\usepackage{bm}% bold math
\usepackage{hyperref}% add hypertext capabilities
\usepackage{IEEEtrantools}

\hypersetup{
   colorlinks=true,
   menucolor=black,
   linkcolor=red,
   citecolor=blue,
   urlcolor=blue}

\makeatletter
\newcommand*{\etal}{%
    \@ifnextchar{.}%
        {\emph{et~al}}%
        {\emph{et~al}.\@\xspace}%
}
\makeatother

\makeatletter
\providecommand*{\diff}%
  {\@ifnextchar^{\DIfF}{\DIfF^{}}}
\def\DIfF^#1{%
  \mathop{\mathrm{\mathstrut d}}%
    \nolimits^{#1}\gobblespace}
\def\gobblespace{%
    \futurelet\diffarg\opspace}
\def\opspace{%
  \let\DiffSpace\!%
  \ifx\diffarg(%
    \let\DiffSpace\relax
  \else
    \ifx\diffarg[%
    \let\DiffSpace\relax
    \else
      \ifx\diffarg\{%
    \let\DiffSpace\relax
      \fi\fi\fi\DiffSpace}
\providecommand*{\Diff}%
  {\@ifnextchar^{\DDIfF}{\DDIfF^{}}}
\def\DDIfF^#1{%
  \mathop{\mathrm{\mathstrut D}}%
    \nolimits^{#1}\gobblespace}
\makeatother

\newcommand{\figref}[1]{Fig.~\ref{fig:#1}}
\newcommand{\Figref}[1]{Figure~\ref{fig:#1}}

\newcommand{\tabref}[1]{Table~\ref{tab:#1}}
\renewcommand{\eqref}[1]{Eq.~(\ref{eq:#1})}

\newcommand{\citeasnoun}[1]{Ref.~\cite{#1}}

\newcommand*\around{{\raise.17ex\hbox{$\scriptstyle\mathtt{\sim}$}}}

\newcommand*\parornament{\vskip +0.8\baselineskip \centerline{*\quad*\quad*} \vskip +0.8\baselineskip}

\newcommand{\keyterm}[1]{\smallskip\noindent\textbf{#1}}
\begin{document}

\title{Principles of Neuromorphic Photonics}% Force line breaks with \\

\author{Bhavin J. Shastri}
    \email{shastri@ieee.org}
\author{Alexander N. Tait}
\author{Thomas Ferreira de Lima}
\author{Mitchell A. Nahmias}
\author{Hsuan-Tung Peng}
\author{Paul R. Prucnal}
\affiliation{Department of Electrical Engineering, Princeton University, Princeton, NJ 08544, USA}%

\date{\today}

%\begin{abstract}
%An article usually includes an abstract, a concise summary of the work covered at length in the main body of the article. Abstract to come here.
%\end{abstract}

\maketitle

%\tableofcontents

\section*{Glossary}\label{sec:glossary}

\keyterm{Benchmark} A standardized task that can be performed by disparate computing approaches, used to assess their relative processing merit in specific cases.

\keyterm{Bifurcation} A qualitative change in behavior of a dynamical system in response to parameter variation. Examples include cusp (from monostable to bistable), Hopf (from stable to oscillating), transcritical (exchange of stability between two steady states).

\keyterm{Brain-inspired computing} (a.k.a. neuro-inspired computing) A biologically inspired approach to build processors, devices, and computing models for applications including adaptive control, machine learning, and cognitive radio. Similarities with biological signal processing include architectural, such as distributed, representational, such as analog or spiking, or algorithmic, such as adaptation.

\keyterm{Broadcast and Weight} A multi-wavelength analog networking protocol in which multiple all photonic neuron outputs are multiplexed and distributed to all neuron inputs. Weights are reconfigured by tunable spectral filters.

\keyterm{Excitability} A far-from-equilibrium nonlinear dynamical mechanism underlying all-or-none responses to small perturbations.

\keyterm{Fan-in} The number of inputs to a neuron.

%\keyterm{Information density} The amount of information flowing through a giving chip area measured in bits per second per square meter.

\keyterm{Layered network} A network topology consisting of a series of sets (i.e., layers) of neurons. The neurons in each set project their outputs only to neurons in the subsequent layer. Most commonly used type of network used for machine learning.

\keyterm{Metric} A quantity assessing performance of a device in reference to a specific computing approach.

\keyterm{Microring weight bank} A silicon photonic implementation of a reconfigurable spectral filter capable of independently setting transmission at multiple carrier wavelengths.

\keyterm{Modulation} The act of representing an abstract variable in a physical quantity, such as photon rate (i.e., optical power), free carrier density (i.e., optical gain), carrier drift (i.e., current). Electrooptic modulators are devices that convert from an electrical signal to the power envelope of an optical signal.

\keyterm{Moore's law} An observation that the number of transistors in an integrated circuit doubles every 18 to 24 months, doubling its performance.

\keyterm{Multiple-Accumulate (MAC)} A common operation that represents a single  multiplication followed by an addition: $a\leftarrow a+(b\times c)$.

\keyterm{Neural networks} A wide class of computing models consisting of a distributed set of nodes, called neurons, interconnected with configurable or adaptable strengths, called weights. Overall neural network behavior can be extremely complex relative to single neuron behavior.

\keyterm{Neuromorphic computing} Computing approaches based on specialized hardware that formally adheres one or more neural network models. Algorithms, metrics, and benchmarks can be shared between disparate neuromorphic computers that adhere to a common mathematical model.

\keyterm{Neuromorphic photonics} An emerging field at the nexus of photonics and neural network processing models, which combines the complementary advantages of optics and electronics to build systems with high efficiency, high interconnectivity, and extremely high bandwidth.

%\keyterm{Spiking} A sparse coding scheme that uses pulses or spikes to encode information in time.recognized by the neuroscience community as an important neural coding strategy for information processing.

\keyterm{Optoelectronics} A technology of electronic devices and systems (semiconductor lasers, photodetectors, modulators, photonic integrated circuits that interact (source, detect and control). %Example of optolectronic devices include.

\keyterm {Photonic integrated circuits (PICs)} A chip that integrates many photonic components (lasers, modulators, filters, detectors) connected by optical waveguides that guide light; similar to an electronic integrated circuit that consists of transistors, diodes, resistors, capacitors, and inductors, connected by conductive wires.

\keyterm{Physical cascadability} The ability of one neuron to produce an output with the same representational properties as its inputs. For example, photonic-electronic-photonic, or 32bit-analog-32bit.

\keyterm{Recurrent network} A network topology in which each neuron output can reach every other neuron, including itself. Every network is a subset of a recurrent network.

\keyterm{Reservoir computing} A computational approach in which a complex, nonlinear substrate performs a diversity of functions, from which linear classifiers extract the most useful information to perform a given algorithm. The reservoir can be implemented by a recurrent neural network or a wide variety of other systems, such as time-delayed feedback.

\keyterm{Semiconductor lasers} Lasers based on semiconductor gain media, where optical gain is achieved by stimulated emission at an interband transition under conditions of a high carrier density in the conduction band.

\keyterm{Signal cascadability} The ability of one neuron to elicit an equivalent or greater response when driving multiple other neurons. The number of target neurons is called fan-out.

\keyterm {Silicon photonics} A chip-scale, silicon on insulator (SOI) platform for monolithic integration of optics and microelectronics for guiding, modulating, amplifying, and detecting light.

\keyterm{Spiking neural networks (SNNs)} A biologically realistic neural network model that processes information with spikes or pulses that encode information temporally.

\keyterm{Wavelength-divison multiplexing (WDM)} One of the most common multiplexing techniques used in optics where different wavelengths (colors) of light are combined, transmitted, and separated again.

\keyterm {Weighted addition} The operation describing how multiple inputs to a neuron are combined into one variable. Can be implemented in the digital domain by multiply-accumulate (MAC) operations or in the analog domain by various physical processes (e.g., current summing, total optical power detection).

\keyterm{WDM weighted addition} A simultaneous summation of power modulated signals and transduction from multiple optical carriers to one electronic carrier. Occurs when multiplexed optical signals impinge on a standard photodetector.

\keyterm{Weight matrix} A way to describe all network connection strengths between neurons arranged such that rows are input neuron indeces and columns are output neuron indeces. The weight matrix can be constrained to be symmetric, block off-diagonal, sparse, etc. to represent particular kinds of neural networks.

%\smallskip\noindent
%\textbf{Dennards law}

\section*{Definition of the Subject}
\label{sec:preface}

In an age overrun with information, the ability to process reams of data has become crucial. The demand for data will continue to grow as smart gadgets multiply and become increasingly integrated into our daily lives. Next-generation industries in artificial intelligence services and high-performance computing are so far supported by microelectronic platforms. These data-intensive enterprises rely on continual improvements in hardware. Their prospects are running up against a stark reality: conventional one-size-fits-all solutions offered by digital electronics can no longer satisfy this need, as Moore's law (exponential hardware scaling), interconnection density, and the von Neumann architecture reach their limits.

With its superior speed and reconfigurability, analog photonics can provide some relief to these problems; however, complex applications of analog photonics have remained largely unexplored due to the absence of a robust photonic integration industry. Recently, the landscape for commercially-manufacturable photonic chips has been changing rapidly and now promises to achieve economies of scale previously enjoyed solely by microelectronics.

Despite the advent of commercially-viable photonic integration platforms, significant challenges still remain before scalable analog photonic processors can be realized.  A central challenge is the development of mathematical bridges linking photonic device physics to models of complex analog information processing.  Among such models, those of neural networks are perhaps the most widely studied and used by machine learning and engineering fields.

\begin{figure*}[!ht]
  \centering
  \includegraphics[width=0.8\linewidth]{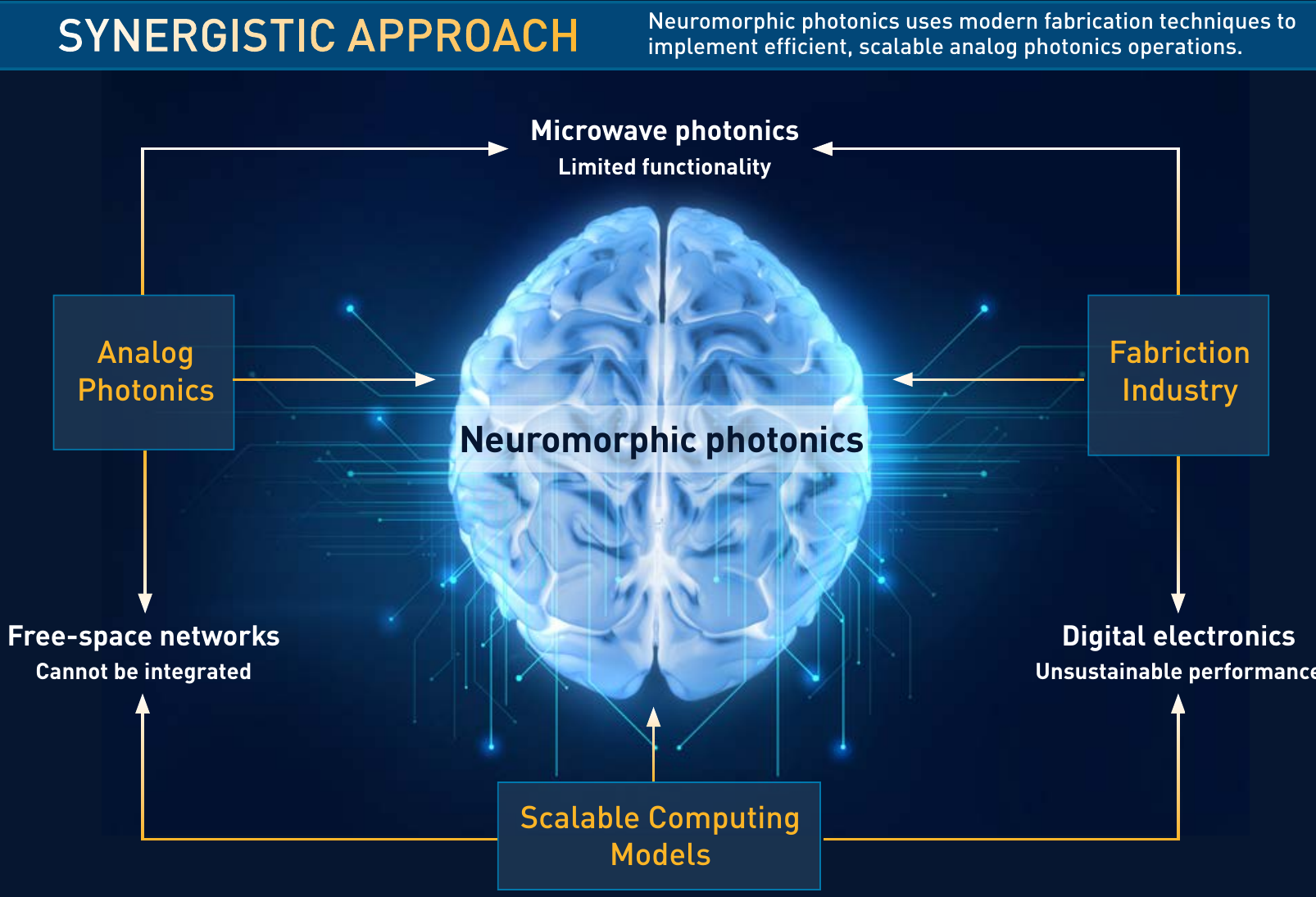}
  \caption{The advent of neuromorphic photonics is due to the convergence of recent advances in photonic integration technology, resurgence of scalable computing models (e.g. spiking, deep neural networks), and a large scale silicon industrial ecosystem.}
  \label{fig:landscape}
\end{figure*}

Recently, the scientific community has set out to build bridges between the domains of photonic device physics and neural networks, giving rise to the field of \emph{neuromorphic photonics} (\figref{landscape}). This article reviews the recent progress in integrated neuromorphic photonics. We provide an overview of neuromorphic computing, discuss the associated technology (microelectronic and photonic) platforms and compare their metric performance. We discuss photonic neural network approaches and challenges for integrated neuromorphic photonic processors while providing an in-depth description of photonic neurons and a candidate interconnection architecture. We conclude with a future outlook of neuro-inspired photonic processing.

%In this article, we reviews the progress in neuromorphic photonics, focusing on photonic integrated devices. The challenges and design rules for optoelectronic instantiation of artificial neurons are presented. The proposed photonic architecture revolves around the processing network node composed of two parts: a nonlinear element and a network interface. We then survey excitable lasers in the recent literature as candidates for the nonlinear node and microring-resonator weight banks as the network interface. Finally, we compare metrics between neuromorphic electronics and neuromorphic photonics and discuss potential applications.

\section*{Introduction}
\label{sec:introduction}

Complexity manifests in our world in countless ways~\cite{Strogatz:2001aa,Vicsek:2002aa} ranging from intracellular processes~\cite{Crescenzi:1998} and human brain area function~\cite{Markram:2011} to climate dynamics~\cite{Donges2009} and world economy~\cite{Hidalgo482}. An understanding of complex systems is a fundamental challenge facing the scientific community. Understanding complexity could impact the progress of our society as a whole, for instance, in fighting diseases, mitigating climate change or creating economic benefits.

Current approaches to complex systems and big-data analysis are based on software, executed on serialized and centralized von Neumann machines.
% Vast volumes of data are synthesized by computational tools into probable models. % and predictive knowledge.
However, the interconnected structure of complex systems (i.e., many elements interacting strongly and with variation~\cite{Bhalla:1999aa}) makes them challenging to reproduce in this conventional computing framework. Memory and data interaction bandwidths constrain the types of informatic systems that are feasible to simulate. The conventional computing approach has persisted thus far due to an exponential performance scaling in digital electronics, most famously embodied in Moore's law. For most of the past 60 years, the density of transistors, clock speed, and power efficiency in microprocessors has approximately doubled every 18 months. These empirical laws are fundamentally unsustainable. The last decade has witnessed a statistically significant ($>$99.95\% likelihood~\cite{Marr:2013}) plateau in processor energy efficiency.

This situation suggests that the time is ripe for radically new approaches to information processing, one being \emph{neuromorphic photonics}. An emerging field at the nexus of photonics and neuroscience, neuromorphic photonics combines the complementary advantages of optics and electronics to build systems with high efficiency, high interconnectivity, and extremely high bandwidth. This article reviews the recent progress in integrated neuromorphic photonics research. First, we introduce neuromorphic computing and give an overview of current technology platforms in microelectronics and photonics. Next, we discuss the evolution of neuromorphic photonics, the present photonic neural network approaches, and challenges for emerging integrated neuromorphic photonic processors. We introduce the concept of a ``photonic neuron'' followed by a discussion on its feasibility. We summarize recent research on optical devices that could be used as network-compatible photonic neurons. We discuss a networking architecture that efficiently channelizes the transmission window of an integrated waveguide. We also compare metrics between neuromorphic electronics and neuromorphic photonics. Finally, we offer a glimpse at the future outlook for neuro-inspired photonic processing and discuss potential applications.

\begin{figure*}[!ht]
  \centering
  \includegraphics[width=1.0\linewidth]{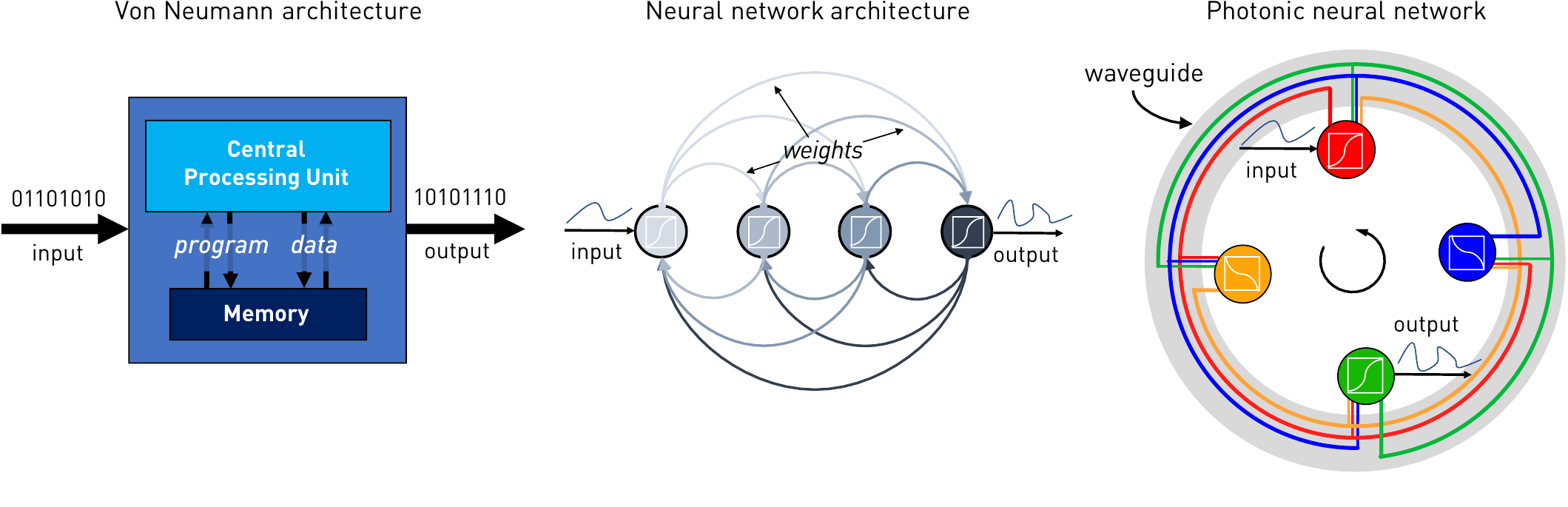}
  \caption{Neural nets: The photonic edge. Von Neumann architectures (left), relying on sequential input-output through a central processor, differ fundamentally from more decentralized neural network architectures (middle). Photonic neural nets (right) can solve the interconnect bottleneck by using one waveguide to carry signals from many connections (easily $N^2\around10,000$) simultaneously.}
  \label{fig:photonic_edge}
\end{figure*}

\section*{Neuromorphic Computing: Beyond von Neumann and Moore}
\label{sec:neuromorphic_computing}

Conventional digital computers are based on the von Neumann architecture~\cite{vonNeumann:1993}. It consists of a memory that stores both data and instructions, a central processing unit (CPU) and inputs and outputs (\figref{photonic_edge} (left)). Instructions and data stored in the memory unit are separated from the CPU by a shared digital bus. This is known as the von Neumann bottleneck~\cite{Backus:1978} which fundamentally limits the performance of the system---a problem that is aggravated as CPUs become faster and memory units larger. This computing paradigm has dominated for over \SI{60}{years} driven in part by the continual progress dictated by Moore's law~\cite{Moore:2000} for CPU scaling and Koomey's law~\cite{Koomey:2011} for energy efficiency (in multiply-accumulate (MAC) operations per joule) compensating the bottleneck. Over the last several years, such scaling has not followed suit, approaching an asymptote. The computation efficiency levels off below \SI{10}{GMAC\per\s\per\W} or \SI{100}{\pico\joule} per MAC~\cite{Hasler2013}. The reasons behind this trend can be traced to both the representation of information at the physical level and the interaction of processing with memory at the architectural level~\cite{Marr:2013}.

Breaching the energy efficiency wall of digital computation by orders of magnitude is not fundamentally impossible. In fact, the human brain, believed to be the most complex system in the universe, is estimated to execute an amazing $\SI{E18}{MAC\per\s}$ using only \SI{20}{\W} of power~\cite{Hasler2013,Merkle:1989}. It does this with \num{E11} neurons with an average of \num{E4} inputs each. This leads to an estimated total of \num{E15} \emph{synaptic} connections, all conveying signals up to \SI{1}{\kHz} bandwidth. The calculated computational efficiency for the brain ($<\si{\atto\J/MAC}$) is therefore 8 orders of magnitude beyond that of current supercomputers ($100\si{\pico\J/MAC}$). The brain is a natural standard for information processing, one that has been compared to artificial processing systems since their earliest inception. Nevertheless, the brain as a processor differs radically from computers today, both at the physical level and at the architectural level. Its exceptional performance is, at least partly, due to the neuron biochemistry, its underlying architecture, and the biophysics of neuronal computation algorithms.

\emph{Neuromorphic computing} offers hope to building large-scale ``bio-inspired'' hardware whose computational efficiencies move toward those of a human brain. In doing so, neuromorphic platforms (\figref{photonic_edge} (middle)) could break performance limitations inherent in traditional von Neumann architectures in solving particular classes of problems. Their distributed hardware architectures can most efficiently evaluate models with high data interconnection, among which are real-time complex system assurance and big data awareness.

At the device level, digital CMOS is reaching physical limits~\cite{Mathur:2002,Taur:1997}. As the CMOS feature sizes scale down below \SIrange{90}{65}{\nm}, the voltage, capacitance, and delay no longer scale according to a well-defined rate by Dennard's law~\cite{Dennard:1974}. This leads to a tradeoff between performance (when transistor is on) and subthreshold leakage (when it is off). For example, as the gate oxide (which serves as an insulator between the gate and channel) is made as thin as possible (\SI{1.2}{\nm}, around five atoms thick Si) to increase the channel conductivity, a quantum mechanical phenomenon of electron tunneling~\cite{Taur:2002,Lee:2001} occurs between the gate and channel leading to increased power consumption. The recent shift to multi-core scaling alleviated these constraints, but the breakdown of Dennard scaling has limited the number of cores than can simultaneously be powered on with a fixed power budget and heat extraction rate. Fixed power budgets have necessitated so called \emph{dark silicon} strategies~\cite{Esmaeilzadeh2012}. Projections for the \SI{8}{\nano\meter} node indicate that over 50\% of the chip will be \emph{dark}~\cite{Esmaeilzadeh2012}, meaning unused at a given time. This has led to a widening rift between conventional computing capabilities and contemporary computing needs, particularly for the analysis of complex systems.

Computational tools have been revolutionary in hypothesis testing and simulation. They have led to the discovery of innumerable theories in science, and they will be an indispensable aspect of a holistic approach to problems in big data and many body physics; however, huge gaps between information structures in observed systems and standard computing architectures motivates a need for alternative paradigms if computational abilities are to be brought to the growing class of problems associated with complex systems.
Brain-inspired computing approaches share the interconnected causal structures and dynamics analogous to the complex systems present in our world. This is in contrast to conventional approaches, where these structures are virtualized at considerable detriment to energy efficiency. From a usual constraint of a fixed power budget, energy efficiency caps overall simulation scale.

Over the years, there has been a deeply committed exploration of unconventional computing techniques~\cite{Keyes:1985,Jaeger02042004,Hasler2013,Merolla08082014,Modha:2011,Tucker:2010aa,Caulfield:2010aa,Woods:2012aa,Benjamin:2014,Pfeil:2013,Furber:2014,Snider:2007,Eliasmith:2012,Indiveri:2011,Brunner:2013aa,Tait:2017,Shen:16arxiv,Shainline:2017,Prucnal:16,PrucnalBook} to alleviate the device level and system/architectural level challenges faced by conventional computing platforms. %Neuromorphic systems allow for simultaneous communication, computation, and memory access throughout the architecture.
Specifically, neuromorphic computing is going through an exciting period as it promises to make processors that use low energies while integrating massive amounts of information. Neuromorphic engineering aims to build machines employing basic nervous systems operations by bridging the physics of biology with engineering platforms enhancing performance for applications interacting with natural environments such as vision and speech~\cite{Hasler2013}. These neural-inspired systems are exemplified by a set of computational principles, including hybrid analog-digital signal representations (discussed next), co-location of memory and processing, unsupervised statistical learning, and distributed representations of information.

\section*{Technology Platforms}
\label{sec:technology_platforms}

%Proposals of unconventional information processing technologies should take into consideration that there is always a cost associated with developing new approaches and new technology. For an information processing technology to be worth adopting, it must sufficiently outperform conventional approaches at a reasonably broad number of problems to outweigh these costs. This section serves as an prelude to Chapter~\ref{chapterComparison} which does a rigorous comparison of various neuromorphic platforms. %Electronic neuromorphic architectures, for example, offer potential orders-of-magnitude performance advantages in energy efficiency. Although they are not universal, they are applicable to a wide range of problems in computer vision, voice recognition, brain simulation, and stuff... Their compatibility with traditional CMOS manufacturing technologies also make them relatively low-cost.

Information representation can have a profound effect on information processing. %In what is considered the third generation of neuromorphic electronics, approaches are typified by their use of \emph{spiking} signals.
 The \emph{spiking} model found in biology is a sparse coding scheme recognized by the neuroscience community as a neural encoding strategy for information processing~\cite{Ostojic:2014aa,paugam2012computing,Kumar:2010aa,Izhikevich:2003,Diesmann:1999aa,Borst:1999aa}, and has code-theoretic justifications~\cite{Sarpeshkar:1998,Thorpe:2001,Maass2002}. Spiking approaches promise extreme improvements in computational power efficiency~\cite{Hasler2013} because they directly exploit the underlying physics of biology~\cite{Jaeger02042004,Maass1997,Sarpeshkar:1998,Izhikevich2007}, analog electronics, or, in the present case, optoelectronics. Digital in amplitude but temporally analog, spiking naturally interleaves robust, discrete representations for communication with precise, continuous representations for computation in order to reap the benefits of both digital and analog. Spiking has two primary advantages over synchronous analog: (1) its analog variable (time) is less noisy than its digital variable (amplitude), and (2) it is asynchronous, without a global clock. Clock synchronization allows for time-division multiplexing (TDM); however, it is a significant practical problem in many-core systems~\cite{Mundy:15}. These advantages may account for the ubiquity of spiking in natural processing systems.~\cite{Thorpe:2001}. It is natural to deepen this distinction to include physical representational aspects, with the important result that optical carrier noise does not accumulate. As will be discussed, when an optical pulse is generated, it is transmitted and routed through a linear optical network with the help of its wavelength identifier.

\begin{figure}[!ht]
  \centering
  \includegraphics[width=1.0\linewidth]{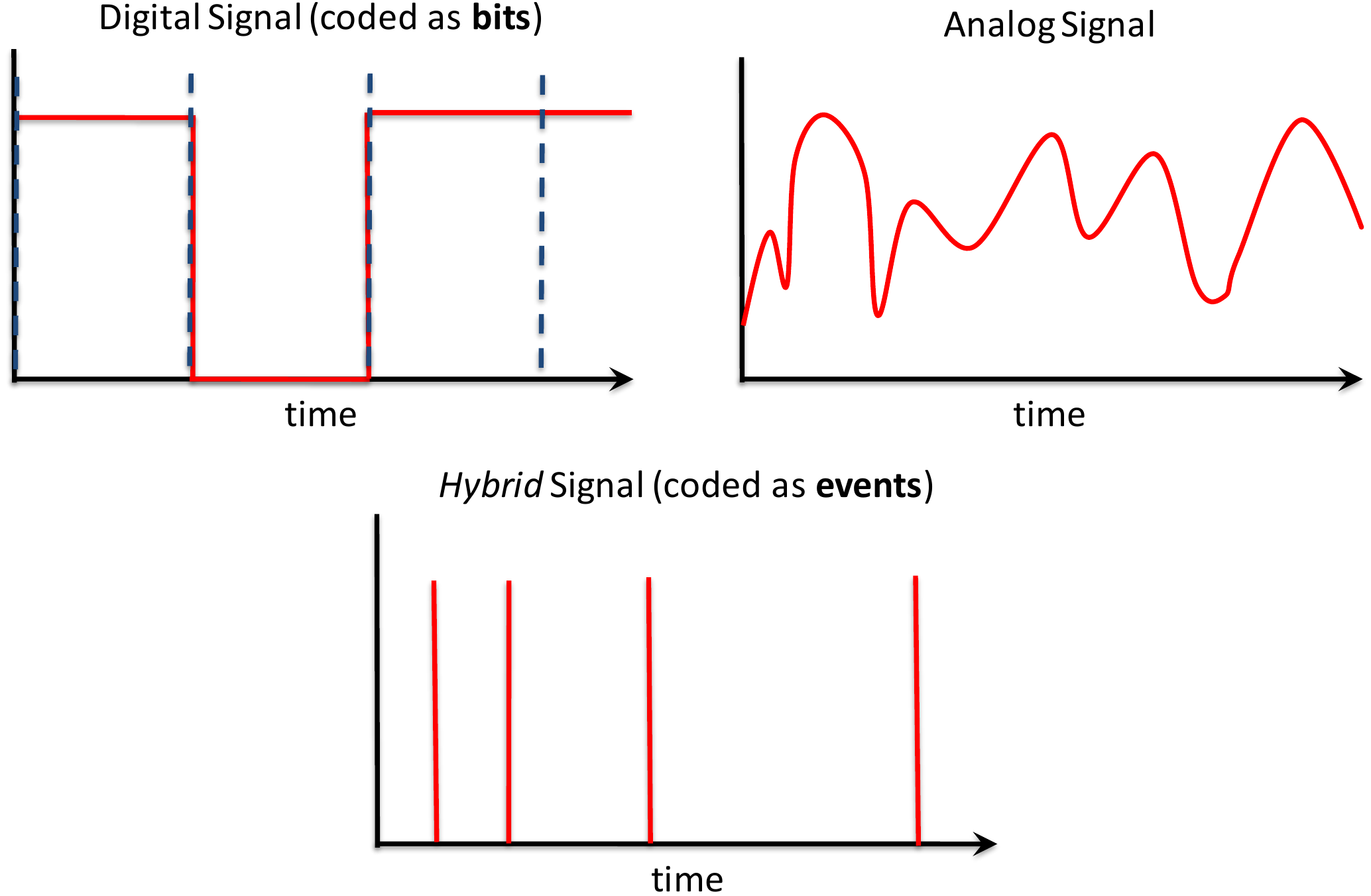}
  \caption{Spiking neural networks encode information as events in time rather than bits. Because the time at which a spike occurs is analog while its amplitude is digital, the signals use a mixed-signal or hybrid encoding scheme. Reproduced from Tait~\emph{et al.} \emph{Photonic Neuromorphic Signal Processing and Computing} ch. 8, 183--222 (2014) in \emph{Nanophotonic Information Physics} by M. Naruse, Ed. Ref~\cite{Tait2014}. With permission of Springer.}
  \label{fig:digital_vs_hybrid}
\end{figure}

%There are a variety of schemes that could be used that are not necessarily spiking. The simplest hybrid analog-digital units are perceptrons, which perform analog computation and use a nonlinear transfer function to suppress analog noise. However, perceptrons lack time-dependent properties and are thus better suited for function approximation rather than the exploration of nonlinear dynamical systems.

\subsection*{Neuromorphic Microelectronics}
\label{subsec:neuromorphic_microelectronics}

\begin{figure*}[!ht]
  \centering
  \includegraphics[width=.75\linewidth]{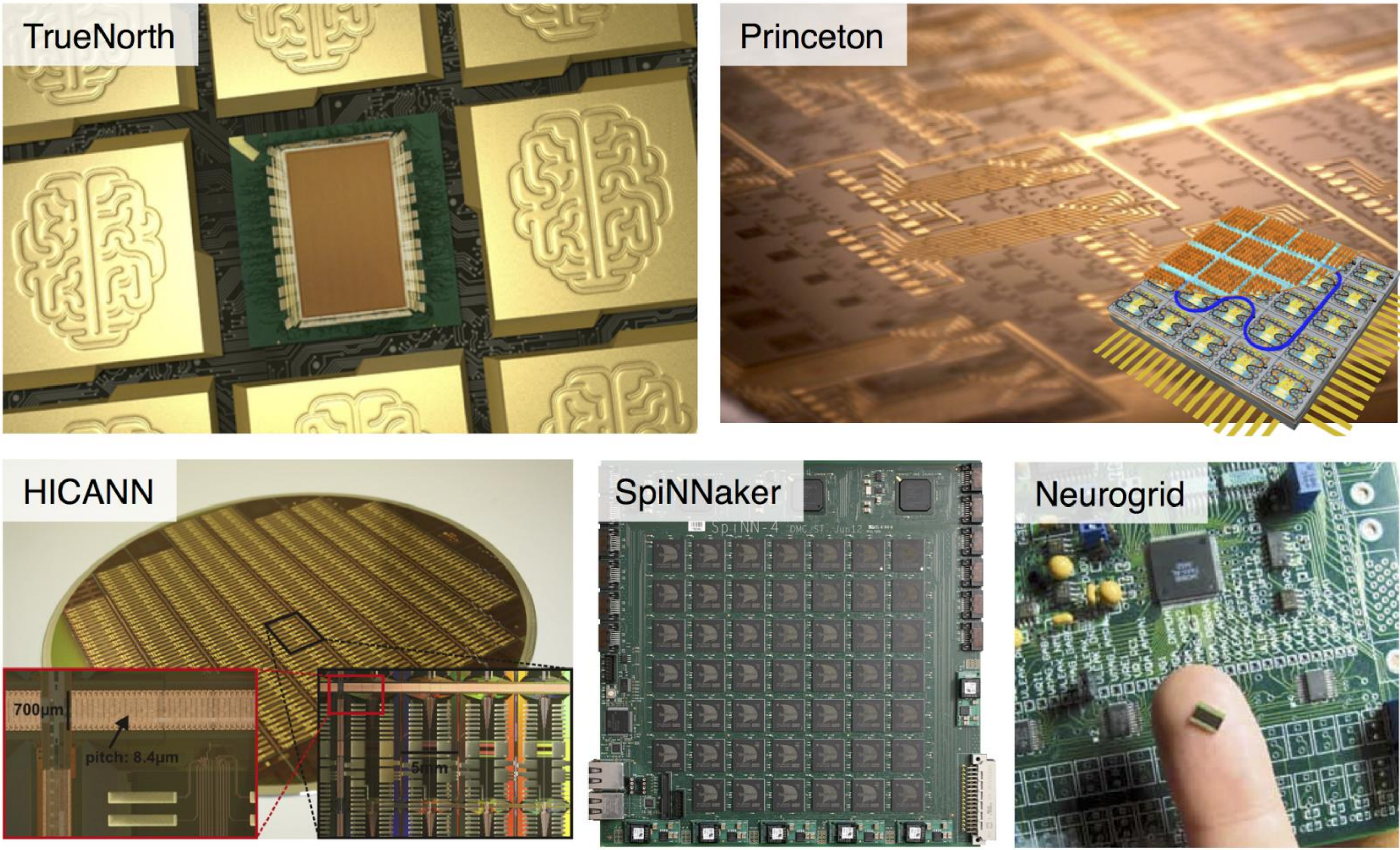}
  \caption{Selected pictures of five different neuromorphic hardware discussed here. They include: TrueNorth~\cite{Merolla08082014}, HICANN~\cite{Schemmel2010}, SpiNNaker~\cite{Furber:2014}, Neurogrid~\cite{Benjamin:2014}.}
  \label{fig:spike_platforms}
\end{figure*}

Spiking primitives have been built in CMOS analog circuits, digital \emph{neurosynaptic cores}, and non-CMOS devices. Various technologies (\figref{spike_platforms}) have demonstrated large-scale spiking neural networks in electronics, including, notably: Neurogrid as part of Stanford University's Brains in Silicon program~\cite{Benjamin:2014}, IBM's TrueNorth as part of DARPA's SyNAPSE program~\cite{Merolla08082014}, HICANN as part of Heidelberg University's FACETS/BrainScaleS project~\cite{Schemmel2010}, and University of Manchester's neuromorphic chip as part of the SpiNNaker project~\cite{Furber:2014}; the latter two are under the flagship of the European Commission's Human Brain Project~\cite{HBP:2012}. These spiking platforms promise potent advantages in efficiency, fault tolerance and adaptability over von Neumann architectures to better interact with natural environments by applying the circuit and system principles of neuronal computation, including robust analog signaling, physics-based dynamics, distributed complexity, and learning. %They promise potent advantages in efficiency, fault tolerance and adaptability over von Neumann architectures for tasks involving machine vision and speech processing.
Using this neuromorphic hardware to process faster signals (e.g., radio waveforms) is, however, not a simple matter of accelerating the clock. These systems rely on slow timescale operation to accomplish dense interconnection.

Whereas von Neumann processors rely on point-to-point memory processor communication, a neuromorphic processor typically requires a large number of interconnects (i.e., \around100s of many-to-one fan-in per processor)~\cite{Hasler2013}. This requires a significant amount of multicasting, which creates a communication burden. This, in turn, introduces fundamental performance challenges that result from RC and radiative physics in electronic links, in addition to the typical bandwidth-distance-energy limits of point-to-point connections~\cite{Miller:2000}. While some incorporate a dense mesh of wires overlaying the semiconductor substrate as crossbar arrays, large-scale systems are ultimately forced to adopt some form of TDM or packet switching, notably, address-event representation (AER), which introduces the overhead of representing spike as digital codes instead of physical pulses. This abstraction at the architectural level allows virtual interconnectivities to exceed wire density by a factor related to the sacrificed  bandwidth, which can be orders of magnitude~\cite{Boahen:2000}. Spiking neural networks (SNNs) based on AER are thus effective at targeting biological time scales and the associated application space: real time applications (object recognition) in the kHz regime~\cite{Merolla08082014,Furber:2014} and accelerated simulation in the low MHz regime~\cite{Schemmel2010}. However, neuromorphic processing for high-bandwidth applications in the GHz regime (such as sensing and manipulating the radio spectrum and for hypersonic aircraft control) must take a fundamentally different approach to interconnection.%, breaking the interconnectivity density and bandwidth tradeoff.

\begin{figure}[!ht]
  \centering
  \includegraphics[width=1.0\linewidth]{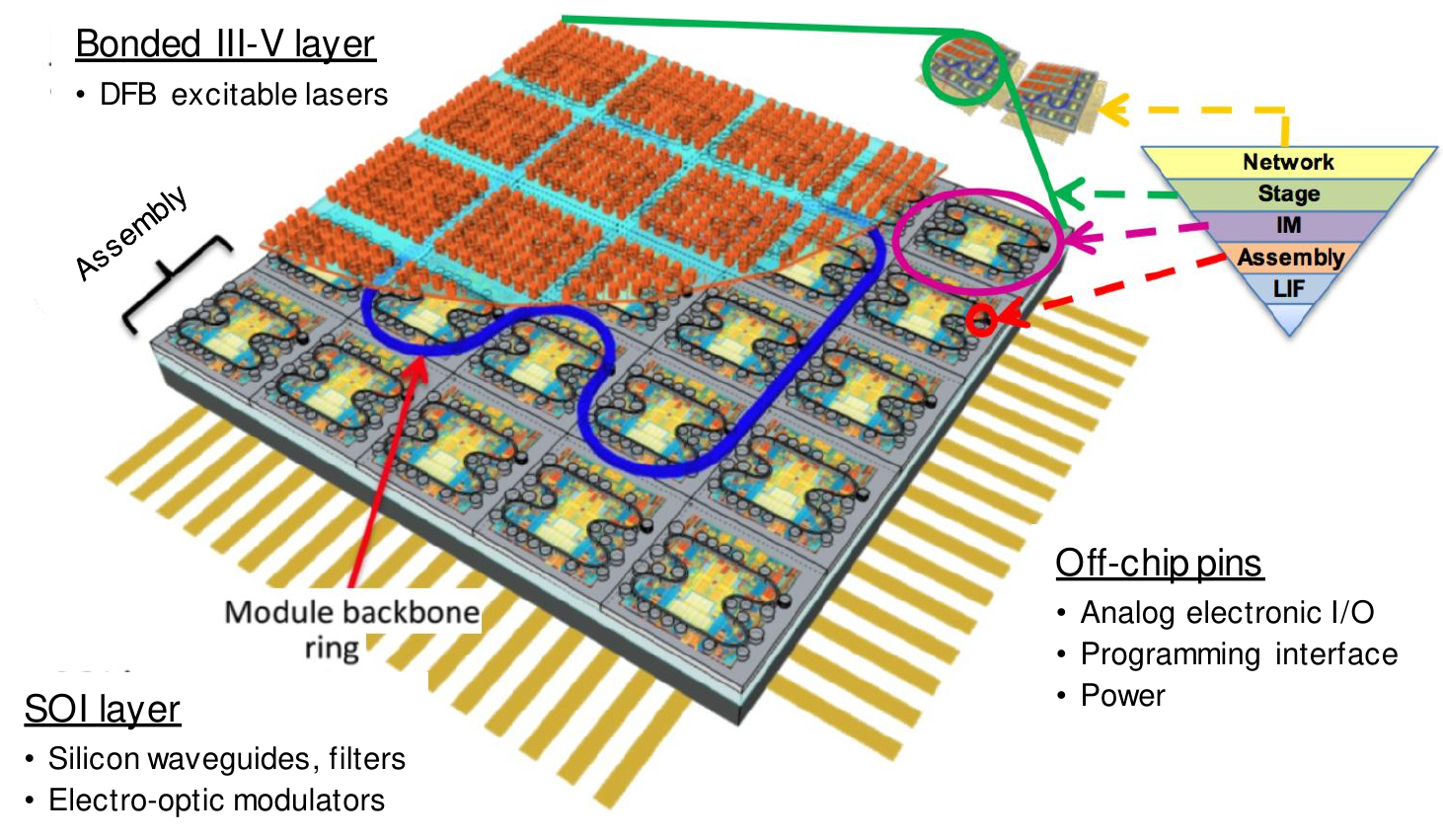}
  \caption{Conceptual rendering of a photonic neuromorphic processor. Laser arrays (orange layer) implement pulsed (spiking) dynamics with electro-optic physics, and a photonic network on-chip (blue and gray) supports complex structures of virtual interconnection amongst these elements while electronic circuitry (yellow) controls stability, self-healing, and learning.}
  \label{fig:laser_brain}
\end{figure}

\subsection*{Toward Neuromorphic Photonics}
\label{subsec:toward_neuromorphic_photonics}

Just as optics and photonics are being employed for interconnection in conventional CPU systems, optical networking principles can be applied to the neuromorphic domain (\figref{photonic_edge} (right)). Mapping a processing paradigm to its underlying dynamics, than abstracting the physics away entirely, can significantly streamline efficiency and performance, and mapping a laser's behavior to a neuron's behavior relies on discovering formal mathematical analogies (i.e., isomorphisms) in their respective governing dynamics. Many of the physical processes underlying photonic devices have been shown to have a strong analogy with biological processing models, which can both be described within the framework of nonlinear dynamics. %Spiking as a model of computation with photonics has shown to be fundamentally enabled by the strong analogy between the underlying physics of biological neuron dynamics and excitable lasers, both of which can be understood within the framework of dynamical systems theory.
Large scale integrated photonic platforms (see \figref{laser_brain}) offer an opportunity for ultrafast neuromorphic processing that complements neuromorphic microelectronics aimed at biological timescales. The high switching speeds, high communication bandwidth, and low crosstalk achievable in photonics are very well suited for an ultrafast spike-based information scheme with high interconnection densities~\cite{Tait:JLT:2014,Prucnal:16}. The efforts in this budding research field aims to synergistically integrate the underlying physics of photonics with spiking neuron-based processing. \emph{Neuromorphic photonics} represents a broad domain of applications where quick, temporally precise and robust systems are necessary. %, including: adaptive control, learning, perception, motion control, sensory processing, autonomous robotics, and cognitive processing of the RF spectrum.

Later in this article we compare metrics between neuromorphic electronics and neuromorphic photonics.

\section*{Neuromorphic Photonics}
\label{sec:neuromorphic_photonics}

Photonics has revolutionized information transmission (communication and interconnects), while electronics, in parallel, has dominated information transformation (computations). This leads naturally to the following question: how can the unifying of the boundaries between the two be made as effective as possible?~\cite{Keyes:1985,Caulfield:2010aa,Tucker:2010aa}. %Research in integrated photonic platforms has followed suit with the rapid development of CMOS-compatible photonic interconnect technologies. This has inadvertently opened a door for unconventional circuit and system opportunities in optics.
CMOS gates only draw energy from the rail when and where called upon; however, the energy required to drive an interconnect from one gate to the next dominates CMOS circuit energy use. Relaying a signal from gate to gate, especially using a clocked scheme, induces penalties in latency and bandwidth compared to an optical waveguide passively carrying multiplexed signals.

This suggests that starting up a new architecture from a photonic interconnection fabric supporting nonlinear optoelectronic devices can be uniquely advantageous in terms of energy efficiency, bandwidth, and latency, sidestepping many of the fundamental tradeoffs in digital and analog electronics. It may be one of the few practical ways to achieve ultrafast, complex on-chip processing without consuming impractical amounts of power~\cite{PrucnalBook}.

\begin{figure}[!ht]
  \centering
  \includegraphics[width=1.0\linewidth]{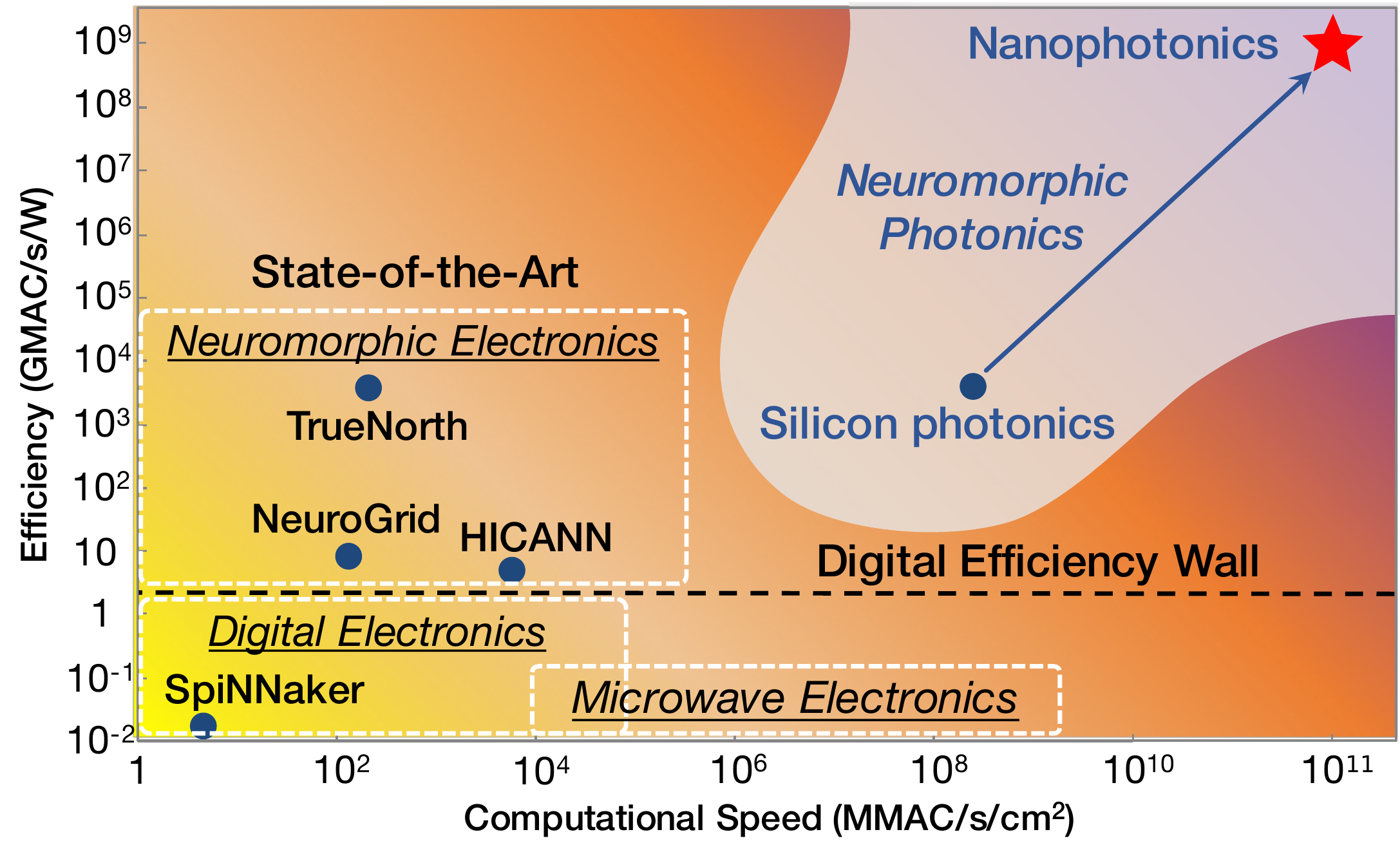}
  \caption{Comparison of neuromorphic hardware platforms. Neuromorphic photonics architectures potentially sport better speed-to-efficiency characteristics than state-of-the-art electronic neural nets (such as IBM’s TrueNorth, Stanford University’s Neurogrid, Heidelberg University's HICANN), as well as advanced digital electronic systems (such as the University of Manchester’s SpiNNaker). On the top-right: the photonic neuron platforms studied in \citeasnoun{PrucnalBook}. The regions highlighted in the graph are approximate, based on qualitative tradeoffs of each technology. Adapted from Ferreira de Lima~\emph{et al.} \emph{Nanophotonics} \textbf{6}, 577--599 (2017) \citeasnoun{FerreiradeLima:2017}. Licensed under Creative Commons Attribution-NonCommercial-NoDerivatives License. (CC BY-NC-ND).}
  \label{fig:mac_comp}
\end{figure}

Complex photonic systems have been largely unexplored due to the absence of a robust photonic integration industry. Recently, however, the landscape for manufacturable photonic chips has been changing rapidly and now promises to achieve economies of scale previously enjoyed solely by microelectronics. In particular, a new photonic manufacturing hybrid platform that combines in the same chip both active (e.g. lasers and detectors), and passive elements (e.g. waveguides, resonators, modulators) is emerging~\cite{Fang:2007,Liang:2010,Liang2010,Roelkens:2010,Heck:2013}. A neuromorphic photonic approach based on this platform could potentially operate 6--8 orders of magnitude faster than neuromorphic electronics when accounting for the bandwidth reduction of virtualizing interconnects~\cite{PrucnalBook} (cf. \figref{mac_comp}; also see \tabref{comparison_neuromorphic} and related discussion).

In the past, the communication potentials of optical interconnects have received attention for neural networking; however, attempts to realize holographic or matrix-vector multiplication systems have failed to outperform mainstream electronics at relevant problems in computing, which can perhaps be attributed to the immaturity of large-scale integration technologies and manufacturing economics.

Techniques in silicon photonic integrated circuit (PIC) fabrication are driven by a tremendous demand for optical interconnects within conventional digital computing systems~\cite{Smit:2012,Jalali2006}, which means platforms for systems integration of active photonics are becoming commercial reality~\cite{Roelkens:2010,Heck:2013,Liang:2010,Liang:2010aa,Marpaung:2013}. %\fxnote{Dest. -A1:}
%Distributed spike processing requires an intimate tie between interconnection and computing at a low level, yet
The potential of recent advances in integrated photonics to enable unconventional computing has not yet been investigated. The theme of current research has been on how modern PIC platforms can topple historic technological barriers between large-scale analog networks and photonic neural systems. In this context, there are two complementary areas of investigation by the research community, namely, photonic spike processing and photonic reservoir computing. While the scope of this article is limited to the former, we briefly introduce both.

\begin{figure}[!ht]
  \centering
  \includegraphics[width=1.0\linewidth]{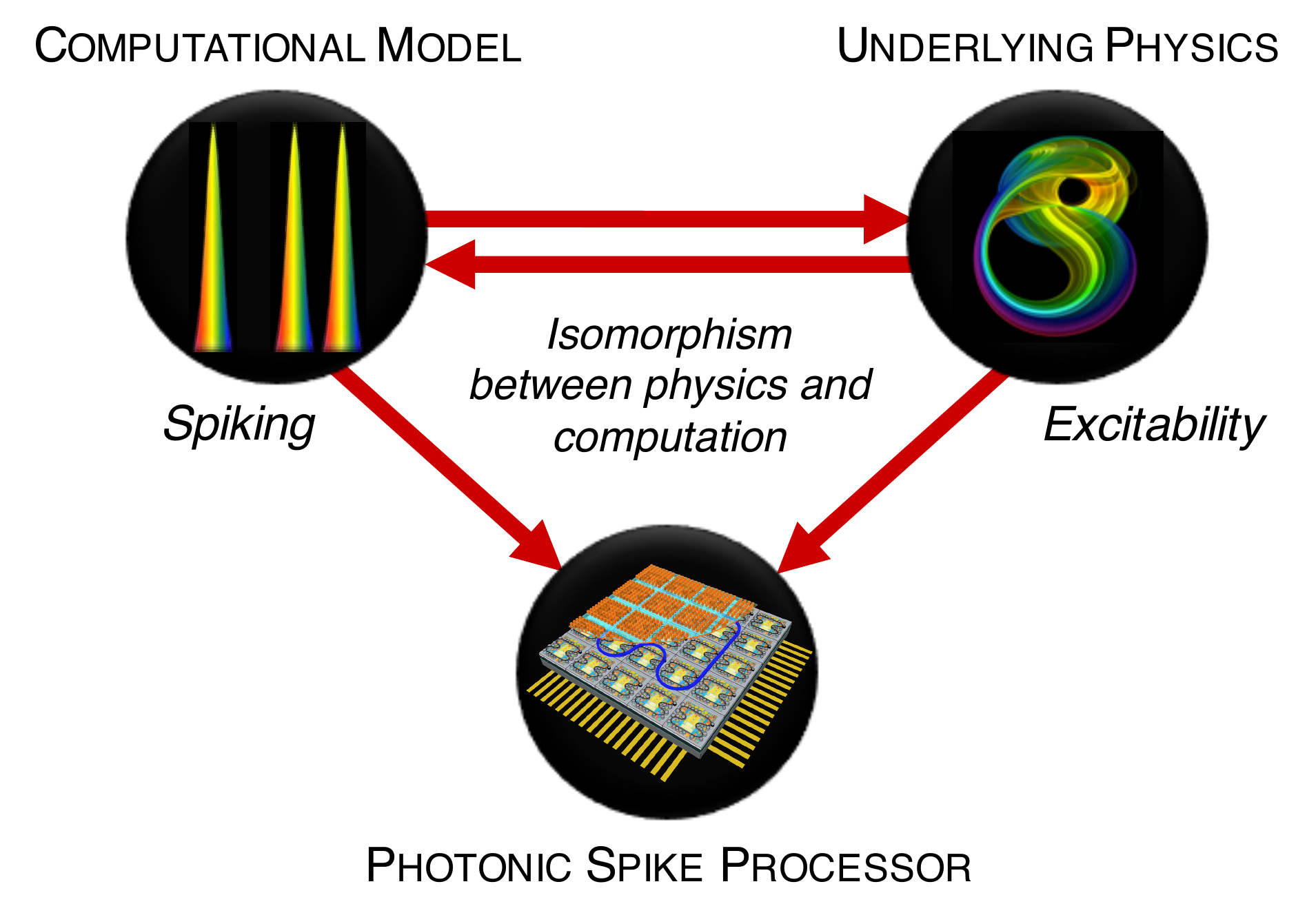}
  \caption{Analogies between spike processing and photonics can be exploited to create a computational paradigm that performs beyond the sum of its parts. By reducing the abstraction between process (spiking) and physics (excitability) there could be a significant advantage on speed, energy usage, scalability. Adapted with permission from Prucnal~\emph{et al.} \emph{Adv. Opt. Photon.} \textbf{8},  228--299 (2016) \citeasnoun{Prucnal:16}. Copyright 2016 Optical Society of America.}
  \label{fig:computing_physics}
\end{figure}

\begin{figure*}[!ht]
  \centering
  \includegraphics[width=0.75\linewidth]{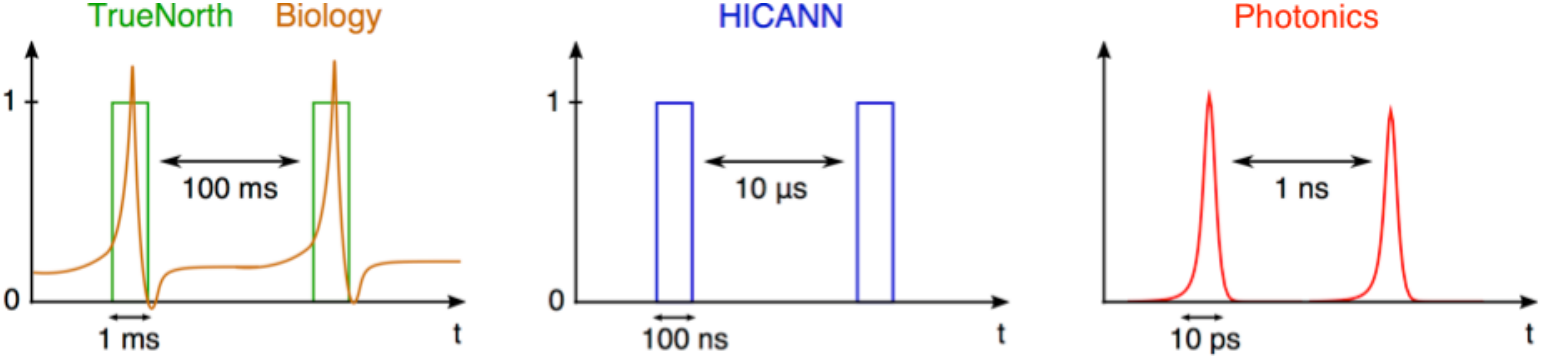}
  \caption{Difference in spike processing time scales (pulse width and refractory period) between biological neurons (left), electronic spiking neurons (middle), and photonic neurons (right). Reproduced with permission from Prucnal~\emph{et al.} \emph{Adv. Opt. Photon.} \textbf{8},  228--299 (2016) \citeasnoun{Prucnal:16}. Copyright 2016 Optical Society of America.}
  \label{fig:time_scales}
\end{figure*}

\subsection*{Photonic Spike Processing}
\label{subsec:photonic_spike_processing}

An investigation of photonics for information processing based on spikes has taken place alongside the development of electronic spiking architectures. Since the first demonstration of photonic spike processing by Rosenbluth~\etal~\cite{Rosenbluth:2009}, there has been a surge in research related to aspects of spike processing in various photonic devices with a recent bloom of proposed forms of spiking dynamics~\cite{Kelleher2010,Kravtsov:2011,Fok:2011,Coomans2011,Brunstein:2012,Nahmias2013,VanVaerenbergh2013,Tait:JLT:2014,Aragoneses:2014,Selmi:2014,Hurtado:APL:2015,Garbin2015,Shastri:2016aa,Romeira:2016aa}---a strategy that could lead to combined computation and communication in the same substrate.

We recently~\cite{Prucnal:16,PrucnalBook} reviewed the recent surge of interest~\cite{Yacomotti:2006,Goulding2007,Kelleher2010,Hurtado2010a,Coomans2011,Brunstein:2012,Coomans2012,VanVaerenbergh2012,Hurtado2012,VanVaerenbergh2013,Nahmias2013,Romeira:2013,Tait2014,Selmi:2014,Aragoneses:2014,Nahmias:2015,Hurtado:APL:2015,Sorrentino:15,Garbin2015,Romeira:2016aa,Shastri:2016aa} in the information processing abilities of semiconductor devices that exploit the dynamical isomorphism between semiconductor photocarriers and neuron biophysics. Many of these proposals for ``photonic neurons'' or ``laser neurons'' or ``optical neurons'' for spike processing are based on lasers operating in an \emph{excitable} regime (\figref{computing_physics}). Excitability~\cite{Hodgkin:1952,Krauskopf2003} is a dynamical system property underlying all-or-none responses. %Excitability~\cite{Hodgkin:1952,Krauskopf2003} is principally a property of far-from-equilibrium nonlinear dynamical systems underlying all-or-none responses to small perturbations that exceed a threshold.

The difference in physical times scales allows these laser systems to exhibit these properties, except many orders of magnitude faster than their biological counterparts~\cite{Nahmias2013}; the temporal resolution (tied to spike widths) and processing speed (tied to refractory period) are accelerated by factors nearing \si{100} million (\figref{time_scales}). %Excitable systems possess unique regenerative properties with strong ties to the underlying physics of devices. %It has been widely recognized as the next significant impact for temporal and spatial information processing\cite{Turconi:2013}.
A network of photonic neurons could open computational domains that demand unprecedented temporal precision, power efficiency, and functional complexity, potentially including applications in wideband radio frequency (RF) processing, adaptive control of multi-antenna systems, and high-performance scientific computing.

\begin{figure}[!ht]
  \centering
  \includegraphics[width=1.0\linewidth]{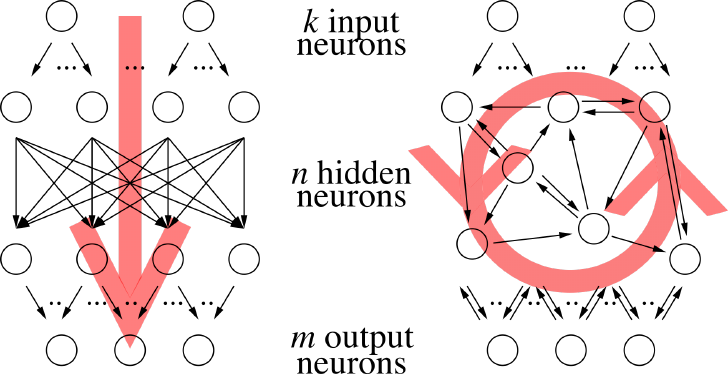}%
  \caption{Comparison of the architectures of a feedforward (left hand side) with a recurrent neural network (right hand side); the grey arrows sketch the possible direction of computation. Adapted from \citeasnoun{Burgsteiner2005b}.}
  \label{fig:feedforward_and_rnn}
\end{figure}

\subsection*{Photonic Reservoir Computing}
\label{subsec:photonic_reservoir_computing}

Reservoir computing (RC) is another promising approach for neuro-inspired computing. %In this framework, a fixed, recurrent network of nonlinear nodes performs a diversity of computations, from which linear classifiers extract the most useful information to perform a given algorithm. These systems maintain many of the advantages of neural networks, including adaptability and robustness to noise. In a hardware context, reservoirs require far fewer tunable elements than traditional neural network models to run effectively. Even simple physical systems can represent more complex virtual networks, and thereby perform a variety of complex tasks. Over the past several years, reservoir computers have been constructed that exploit the incredible bandwidths and speeds available to photonic signals. These `photonic reservoirs' utilized optical multiplexing strategies to form highly complex virtual networks. Experimentally demonstrated systems have displayed state-of-the-art performance in a variety of areas, including speech recognition, time-series prediction, Boolean logic operations, and nonlinear channel equalization.
A central tenet of RC is that complex processes are generated in a medium whose behavior is not necessarily understood theoretically. Instead, the ``reservoir'' (a fixed, recurrent network of nonlinear nodes; see \figref{feedforward_and_rnn}) generates a large number of complex processes, and linear combinations of reservoir signals are trained to approximate a desired task~\cite{Verstraeten2007}. To arrive at a user-defined behavior, reservoirs do not need to model or program and instead rely on supervised machine learning techniques for simple linear classifiers. This is advantageous in systems whose overall behavior is complex, yet difficult to model or correspond to a theoretical behavior. There are a wide class of physical systems that fit that description, and the reservoir concept makes them highly likely to apply to widespread information processing tasks~\cite{Soriano2015}.

%This blindness to internal theory can also be viewed as a tradeoff for RC. A much larger reservoir is needed to achieve an equal level of accuracy to do a given task that can also be implemented through programming or through plasticity within the reservoir. This is due in part to the tremendous growth of generated processes as the reservoir increases in size. In a sense, the efficiency of such an approach becomes worse with the complexity of the desired task because the number of processes that are generated and then unused grow combinatorially, while the number that are used stays constant. The combinatorial growth of side-effect processes can also complicate the procedure to find the correct process through supervised learning. The dimensionality of the output feature vector increases with the number of reservoir variables, and dimensionality can be a key limiter in the convergence rate of known machine learning methods.

Over the past several years, reservoir computers have been constructed that exploit the incredible bandwidths and speeds available to photonic signals. These `photonic reservoirs' utilize optical multiplexing strategies to form highly complex virtual networks. %Photonics-based hardware implementations of RC simultaneously exploit the biophysics of neuronal computational algorithms with the advantages optics offers. (high-bandwidth, high switching speeds, low power consumption, low crosstalk).
Photonic RC particularly is attractive when the information to be processed is already in the optical domain, for example, applications in telecommunications and image processing. Recently, there has been a significant development in the hardware realization of RC. Optical reservoirs have been demonstrated with various schemes such as benchtop demonstrations with a fiber with a single nonlinear dynamical node~\cite{Paquot:2012,Martinenghi2012,Larger:12,Duport2012,Brunner:2013aa,Brunner2013,Hicke2013,Soriano:13,Ortin:2015aa,Duport:2016}, and integrated solutions including microring resonators~\cite{Mesaritakis2013}, a network of coupled semiconductor optical amplifiers (SOAs)~\cite{Vandoorne2011}, and a passive silicon photonics chip~\cite{Vandoorne:2014aa}. %In general, integrated photonic solutions offer advantages over fiber-based solutions in terms of small footprint and mechanical stability, allowing for scalability to large reservoirs.

It has been experimentally demonstrated and verified that some of these photonic RC solutions achieve highly competitive figures of merit at unprecedented data rates often outperforming software-based machine learning techniques for computationally hard tasks such as spoken digit and speaker recognition, chaotic time-series prediction, signal classification, or dynamical system modeling. Another significant advantage of photonic-based approaches, as pointed out by Vandoorne~\etal~\cite{Vandoorne:2014aa}, is the straightforward use of coherent light to exploit both the phase and amplitude of light. The simultaneous exploitation of two physical quantities results in a notable improvement over real-valued networks that are traditionally used in software-based RC---a reservoir operating on complex numbers in essence doubles the internal degrees of freedom in the system, leading to a reservoir size that is roughly twice as large as the same device operated with incoherent light.

%As pointed out by Brunner~\etal~\cite{Brunner:2013aa}, the performance of photonic RC hardware is generally evaluated by focusing on two different, general classes of information processing tasks. The first type involves classification of information, that is, associating different inputs to different classes. These tasks require discrete classes as classifier targets, and a system response sufficiently diverse to allow for clear separation. The second type of tasks is based on nonlinear processing of dynamical information where the classifier target values can be continuous and the system has to provide memory to capture the dynamical nature of information. The above listed information processing tasks fall in either of these categories.

\parornament

Neuromorphic spike processing and reservoir approaches differ fundamentally and possess complementary advantages. Both derive a broad repertoire of behaviors (often referred to as complexity) from a large number of physical degrees-of-freedom (e.g., intensities) coupled through interaction parameters (e.g., transmissions). Both offer means of selecting a specific, desired behavior from this repertoire using controllable parameters. In neuromorphic systems, network weights are \emph{both} the interaction and controllable parameters, whereas, in reservoir computers, these two groups of parameters are separate. This distinction has two major implications. Firstly, the interaction parameters of a reservoir do not need to be observable or even repeatable from system-to-system. Reservoirs can thus derive complexity from physical processes that are difficult to model or reproduce. Furthermore, they do not require significant hardware to control the state of the reservoir. Neuromorphic hardware has a burden to correspond physical parameters (e.g. drive voltages) to model parameters (e.g. weights). Secondly, reservoir computers can only be made to elicit a desired behavior through instance-specific supervised training, whereas neuromorphic computers can be programmed \emph{a priori} using a known set of weights. Because neuromorphic behavior is determined only by controllable parameters, these parameters can be mapped directly between different system instances, different types of neuromorphic systems, and simulations. Neuromorphic hardware can leverage existing algorithms (e.g., Neural Engineering Framework (NEF)~\cite{Stewart:2014}), map virtual training results to hardware, and particular behaviors are guaranteed to occur. Photonic RCs can of course be simulated; however, they have no corresponding guarantee that a particular hardware instance will reproduce a simulated behavior or that training will be able to converge to this behavior.

\subsection*{Challenges for Integrated Neuromorphic Photonics}
\label{sec:challenges}

Key criteria for nonlinear elements to enable a scalable computing platform include: thresholding, fan-in, and cascadability~\cite{Keyes:1985,Caulfield:2010aa,Tucker:2010aa}. Past approaches to optical computing have met challenges realizing these criteria. A variety of digital logic gates in photonics that suppress amplitude noise accumulation have been claimed, but many proposed optical logic devices do not meet necessary conditions of cascadability. Analog photonic processing has found application in high bandwidth filtering of microwave signals~\cite{Capmany2006}, but the accumulation of phase noise, in addition to amplitude noise, limits the ultimate size and complexity of such systems.

Recent investigations~\cite{PrucnalBook,Coomans2011,Brunstein:2012,Nahmias2013,VanVaerenbergh2013,Tait:JLT:2014,Aragoneses:2014,Selmi:2014,Hurtado:APL:2015,Garbin2015,Nahmias:2015,Shastri:2016aa,Romeira:2016aa} have suggested that an alternative approach to exploit the high bandwidth of photonic devices for computing lies not in increasing device performance or fabrication, but instead in examining models of computation that hybridize techniques from analog and digital processing. These investigations have concluded that a photonic neuromorphic processor could satisfy them by implementing a model of a neuron, i.e., a photonic neuron, as opposed to the model of a logic gate. %In addition, schemes that take advantage of the multiple available wavelengths require ubiquitous wavelength conversion, which can be costly, noisy, and inefficient.
Early work in neuromorphic photonics involved fiber-based spiking approaches for learning, pattern recognition, and feedback~\cite{Rosenbluth:2009,Fok:2011,Kravtsov:2011}. Spiking behavior resulted from a combination of SOA together with a highly nonlinear fiber thresholder, but they were neither excitable nor asynchronous and therefore not suitable for scalable, distributed processing in networks.

``Neuromorphism'' implies a strict isomorphism between artificial neural networks and optoelectronic devices. There are two research challenges necessary to establish this isomorphism: the nonlinearity (equivalent to thresholding) in individual neurons, and the synaptic interconnection (related to fan-in and cascadability) between different neurons, as will be discussed in the proceeding sections. Once the isomorphism is established and large networks are fabricated, we anticipate that the computational neuroscience and software engineering will have a new optimized processor for which they can adapt their methods and algorithms.

%Recent investigations have concluded that a photonic unconventional computing primitive called the processing network node (PNN) could satisfy them by implementing a model of a neuron as opposed to the model of a logic gate.

Photonic neurons address the traditional problem of noise accumulation by interleaving physical representations of information. Representational interleaving, in which a signal is repeatedly transformed between coding schemes (digital-analog) or physical variables (electronic- optical), can grant many advantages to computation and noise properties. From an engineering standpoint, the logical function of a nonlinear neuron can be thought of as increasing signal-to-noise ratio (SNR) that tends to degrade in linear systems, whether that means a continuous nonlinear transfer function suppressing analog noise or spiking dynamics curtailing pulse attenuation and spreading. As a result, we neglect purely linear PNNs as they do not offer mechanisms to maintain signal fidelity in a large network in the presence of noise.

The optical channel is highly expressive and correspondingly very sensitive to phase and frequency noise. For example, the networking architecture proposed in Tait~\etal \citeasnoun{Tait:JLT:2014} relies on wavelength-division multiplexing (WDM) for interconnecting many many points in a photonic substrate together. Any proposal for networking computational primitives must address the issue of practical cascadability: transferring information and energy in the optical domain from one neuron to many others and exciting them with the same strength without being sensitive to noise. This is notably achieved, for example, by encoding information in energy pulses that can trigger stereotypical excitation in other neurons regardless of their analog amplitude. In addition, as will be discussed next, schemes which exhibit limitations with regards to wavelength channels may require a large number of wavelength conversion steps, which can be costly, noisy and inefficient.

%Photonic unconventional computing primitives address the traditional problem of noise accumulation by interleaving physical representations of information. Representational interleaving, in which a signal is repeatedly transformed between coding schemes (digital-analog) or physical variables (electronic-optical), can grant many advantages to computation and noise properties. As noted by Sarpeshkar~\cite{Sarpeshkar:1998}, hybrid analog-digital systems---particularly those composed of moderately precise analog units coupled together---maximize processing while minimizing time, energy, and material costs.

\section*{Photonic Neuron}
\label{sec:photonic_neuron}

\subsection*{What is an Artificial Neuron?}
\label{subsec:artificial_neuron}

Neuroscientists research artificial neural networks as an attempt to mimic the \emph{natural processing} capabilities of the brain. These networks of simple nonlinear nodes can be taught (rather than programmed) and reconfigured to best execute a desired task; this is called \emph{learning}. Today, neural nets offer state-of-the-art algorithms for machine intelligence such as speech recognition, natural language processing, and machine vision~\cite{Bengio2013}.

\begin{figure}[!ht]
  \centering
  \includegraphics[width=1.0\linewidth]{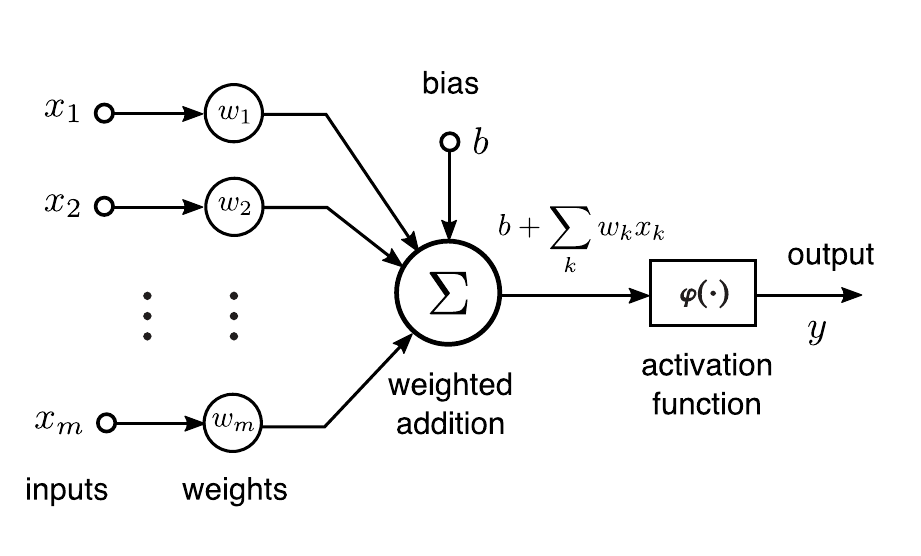}
  \caption{Nonlinear model of a neuron. Note the three parts: (i) a set of \emph{synapses}, or \emph{connecting links}; (ii) an \emph{adder}, or \emph{linear combiner}, performing weighted addition; and (iii) a nonlinear \emph{activation function}. Reproduced with permission from Prucnal~\emph{et al.} \emph{Adv. Opt. Photon.} \textbf{8},  228--299 (2016) \citeasnoun{Prucnal:16}. Copyright 2016 Optical Society of America.}
  \label{fig:nonlinear_neuron_model}
\end{figure}

Three elements constitute a neural network: a set of nonlinear nodes (neurons), configurable interconnection (network), and information representation (coding scheme). An elementary illustration of a neuron is shown in \figref{nonlinear_neuron_model}. The network consists of a weighted directed graph, in which connections are called synapses. The input of a neuron is a linear combination (or weighted addition) of the outputs of the neurons connected to it. Then, the particular neuron integrates the combined signal and produces a nonlinear response, represented by an \emph{activation function}, usually monotonic and bounded.

Three generations of neural networks were historically studied in computational neuroscience~\cite{Maass1997}. The first was based on the McCulloch--Pitts neural model, which consists of a linear combiner followed by a step-like activation function (binary output). These neural networks are boolean-complete---i.e. they have the ability of simulating any boolean circuit and are said to be universal for digital computations. The second generation implemented analog outputs, with a continuous activation function instead of a hard thresholder. Neural networks of the second generation are universal for analog computations in the sense they can uniformly approximate arbitrarily well any continuous function with a compact domain~\cite{Maass1997}. When augmented with the notion of `time', recurrent connections can be created and be exploited to create attractor states~\cite{Eliasmith2005} and associative memory~\cite{Hopfield01041982} in the network.

Physiological neurons communicate with each other using pulses called action potentials or spikes. In traditional neural network models, an analog variable is used to represent the firing rate of these spikes. This coding scheme called \emph{rate coding} was believed to be a major, if not the only, coding scheme used in biology. Surprisingly, there are some fast analog computations in the visual cortex that cannot possibly be explained by rate coding. For example, neuroscientists demonstrated in the 1990s that a single cortical area in macaque monkey is capable of analyzing and classifying visual patterns in just \SI{30}{\ms}, in spite of the fact that these neurons' firing rates are usually below \SI{100}{\Hz}---i.e. less than 3 spikes in \SI{30}{\ms}~\cite{Perrett1982,Maass1997,Thorpe:2001} which directly challenges the assumptions of rate coding. In parallel, more evidence was found that biological neurons use the precise timing of these spikes to encode information, which led to the investigation of a third generation of neural networks based on a \emph{spiking neuron}.

The simplicity of the models of the previous generations precluded the investigation of the possibilities of using \emph{time} as resource for computation and communication. If the \emph{timing} of individual spikes itself carry analog information (\emph{temporal coding}), then the energy necessary to create such spike is optimally employed to express information. Furthermore, Maass~\etal showed that this third generation is a generalization of the first two, and, for several concrete examples, can emulate real-valued neural network models while being more robust to noise~\cite{Maass1997}.

For example, one of the simplest models of a spiking neuron is called \emph{leaky integrate-and-fire} (LIF), described in Eq.~\ref{eq:lif_model}. It represents a simplified circuit model of the membrane potential of a biological spiking neuron.

\begin{IEEEeqnarray}{rCl}
  C_m\frac{\diff{V_m(t)}}{\diff{t}}&&=-\frac{1}{R_m}(V_m(t)-V_L)+I_{app}(t); \label{eq:lif_model}\\
  &&\textrm{if $V_m(t)>V_{\mathrm{thresh}}$ then} \nonumber \\
  &&\textrm{release a spike and set $V_m(t)\rightarrow V_{\mathrm{reset}}$,}\nonumber
\end{IEEEeqnarray}

where $V_m(t)$ is the membrane voltage, $R_m$ the membrane resistance, $V_L$ the equilibrium potential, and $I_{\mathrm{app}}$ the applied current (input). More bio-realistic models, such as the Hodgkin--Huxley model, involve a several ordinary differential equations and nonlinear functions.

However, simply simulating neural networks on a conventional computer, be it of any generation, is costly because of the fundamentally serial nature of CPU architectures. Bio-realistic SNN present a particular challenge because of the need for fine-grained time discretization~\cite{Izhikevich2004}. Engineers circumvent this challenge by employing an event-driven simulation model which resolves this issue by storing the time and shape of the events expanded in a suitable basis in a simulation queue. Although simplified models do not faithfully reproduce key properties of cortical spiking neurons, it allows for large-scale simulations of SNNs, from which key networking properties can be extracted.

Alternatively, one can build an unconventional, distributed network of nonlinear nodes, which directly use the physics of nonlinear devices or excitable dynamical systems, significantly dropping energetic cost per bit.

Here, we discuss recent advances in neuromorphic photonic hardware and the constraints to which particular implementations must subject, including accuracy, noise, cascadability and thresholding. A successful architecture must tolerate eventual inaccuracies and noise, indefinite propagation of signals, and provide mechanisms to counteract noise accumulation as the signal traverses across the network.

\subsection*{Basic Requirements for a Photonic Neuron}
\label{subsec:pnn_requirements}

An artificial neuron described in \figref{nonlinear_neuron_model} must perform three basic mathematical operations: array multiplication (weighting), summation, and a nonlinear transformation (activation function). Moreover, the inputs to be weighted in the first stage must be of the same nature of the output---in the case considered here, photons.

As the size of the network grows, additional mechanisms are required at the hardware level to ensure the integrity of the signals. The neuron must have a scalable number of inputs, referred to as \emph{maximum fan-in} ($N_f$), which will determine the degree of connectivity of the network. Each neuron's output power must be strong enough to drive at least $N_f$ others (\emph{cascadability}). This concept is tied closely with that of \emph{thresholding:} the SNR at the output must be higher than at its input. Cascadability, thresholding, and fan-in are particularly challenging to optical systems due to quantum efficiency (photons have finite supply) and amplified spontaneous emission (ASE) noise, which degrades SNR.

\subsection*{The Processing Network Node}
\label{subsec:pnn_module}

A networkable photonic device with optical I/O, provided that is capable of emulating an artificial neuron, is named a processing-network node (PNN)~\cite{Tait:JLT:2014}. Formulations of a photonic PNN can be divided into two main categories: all-optical and optical-electrical-optical (O/E/O), respectively classified according to whether the information is always embedded in the optical domain or switches from optical to electrical and back. We note that the term \emph{all-optical} is sometimes very loosely defined in engineering articles. Physicists reserve it for devices that rely on parametric nonlinear processes, such as four-wave mixing. Here, our definition includes devices that undergo nonparametric processes as well, such as semiconductor lasers with optical feedback, in which optical pulses directly perturb the carrier population, triggering quick energy exchanges with the cavity field that results in the release of another optical pulse.

Silicon waveguides have a relatively enormous transparency window of \SI{7.5}{THz}~\cite{Agrawal:02} over which they can guide lightwaves with very low attenuation and crosstalk, in contrast with electrical wires or radio frequency transmission lines. With WDM, each input signal exists at a different wavelength but is superimposed with other signals onto the same waveguide. For example, to maximize the information throughput, a single waveguide could carry hundreds of wideband signals (\around\SI{20}{GHz}) simultaneously. As such, it is highly desirable and crucial to design a PNN that is compatible with WDM. All-optical versions of a PNN must have some way to sum multiwavelength signals, and this requires a population of charge carriers. On the other hand, O/E/O versions could make use of photodetectors (PD) to provide a spatial sum of WDM signals. The PD output could drive an E/O converter, involving a laser or a modulator, whose optical output is a nonlinear result of the electrical input. Instances of both techniques are presented in the next section.

\begin{figure}[!ht]
  \centering
  \includegraphics[width=1.0\linewidth]{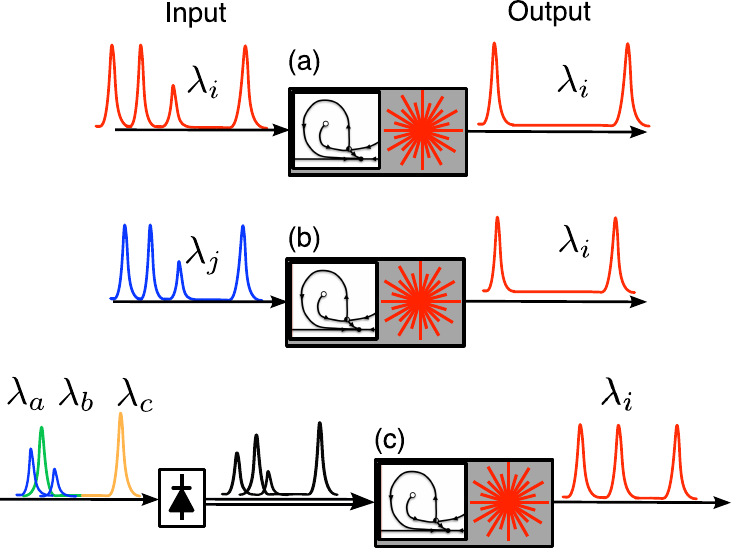}
  \caption{General classification of semiconductor excitable lasers based on: (a) coherent optical injection, (b) non-coherent optical injection and (c) full electrical injection. Each of these lasers can be pumped either electrically or optically. Reproduced from Ferreira de Lima~\emph{et al.} \emph{Nanophotonics} \textbf{6}, 577--599 (2017) \citeasnoun{FerreiradeLima:2017}. Licensed under Creative Commons Attribution-NonCommercial-NoDerivatives License. (CC BY-NC-ND).}
  \label{fig:excitable_lasers}
\end{figure}

\begin{figure*}[!ht]
  \centering
  \includegraphics[width=1.0\linewidth]{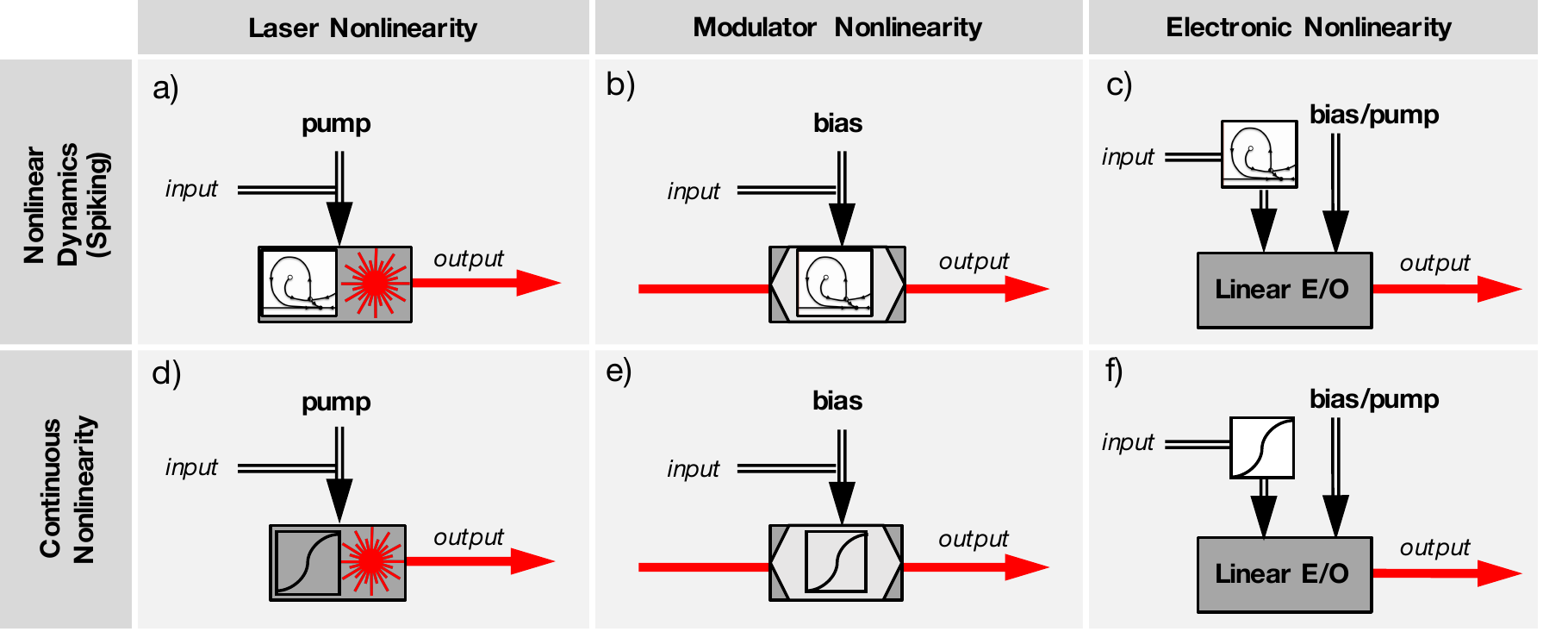}
  \caption{Classification of O/E/O PNN nonlinearities and possible implementations. (a) Spiking laser neuron. (b) Spiking modulator. (c) Spiking or arbitrary electronic system driving a linear electro-optic (E/O) transducer---either modulator or laser. (d) Overdriven continuous laser neuron, as demonstrated in~\cite{Nahmias:2016}. (e) Continuous modulator neuron, as demonstrated in \citeasnoun{Tait:2017}. (f) Continuous purely electronic nonlinearity with optical output. Reproduced from Ferreira de Lima~\emph{et al.} \emph{Nanophotonics} \textbf{6}, 577--599 (2017) \citeasnoun{FerreiradeLima:2017}. Licensed under Creative Commons Attribution-NonCommercial-NoDerivatives License. (CC BY-NC-ND).}
  \label{fig:alt_nodes}
\end{figure*}

\subsubsection*{All-optical PNNs}
\label{subsubsec:all_optical_pnns}
Coherent injection models are characterized by input signals directly interacting with cavity modes, such that outputs are at the same wavelength as inputs (\figref{excitable_lasers}(a)). Since coherent optical systems operate at a single wavelength $\lambda$, the signals lack distinguishability from one another in a WDM-encoded framework. As demonstrated in \citeasnoun{Alexander2013}, the effective weight of coherently injected inputs is also strongly phase dependent. Global optical phase control presents a challenge in synchronized laser systems, but also affords an extra degree of freedom to configure weight values.

Incoherent injection models inject light in a wavelength $\lambda_{j}$ to selectively modulate an intra-cavity property that then triggers excitable output pulses in an output wavelength $\lambda_{i}$ (\figref{excitable_lasers}(b)). A number of approaches~\cite{Selmi:15,Selmi:2014,Nahmias2013,Hurtado:APL:2015}---including those based on optical pumping---fall under this category. While distinct, the output wavelength often has a stringent relationship with the input wavelength. For example, excitable micropillar lasers~\cite{Selmi:2014,Barbay:2011} are carefully designed to support one input mode with a node coincident with an anti-node of the lasing mode. In cases where the input is also used as a pump~\cite{Shastri:2016aa}, the input wavelength must be shorter than that of the output in order to achieve carrier population inversion.

WDM networking introduces wavelength constraints that conflict with the ones inherent to optical-injection. One approach for networking optically-injected devices is to attempt to separate these wavelength constraints. In early work on neuromorphic photonics in fiber, this was accomplished with charge-carrier-mediated cross-gain modulation (XGM) in an SOA~\cite{Rosenbluth:2009,Fok:2011,Kravtsov:2011}.

\subsubsection*{O/E/O PNNs}
\label{subsubsec:oeo_pnns}
In this kind of PNN, the O/E subcircuit is responsible for the weighted addition functionality, whereas the E/O is responsible for the nonlinearity (\figref{excitable_lasers}(c)). Each subcircuit can therefore be analyzed independently. The analysis of an O/E WDM weighted addition circuit is deferred to a later section (photonic neural networks).

The E/O subcircuit of the PNN must take an electronic input representing the complementary weighted sum of optical inputs, perform some dynamical or nonlinear process, and generate a clean optical output on a single wavelength. \Figref{alt_nodes} classifies six different ways nonlinearities can be implemented in an E/O circuit. The type of nonlinearity, corresponding to different neural models, is separated into \emph{dynamical systems} and \emph{continuous nonlinearities}, both of which have a single input $u$ and output $y$. A continuous nonlinearity is described by a differential equation $\dot{y} = f(y,u)$. This includes continuous-time recurrent neural networks (CTRNNs) such as Hopfield networks. The derivative of $y$ introduces a sense of time, which is required to consider recurrent networking, although it does not exclude feedforward models where time plays no role, such as perceptron models. A dynamical system has an internal state $\vec{x}$ and is described by $\dot{\vec{x}} = g(\vec{x},u); \dot{y} = h(\vec{x},y,u)$, where the second differential equation represents the mapping between the internal state $\vec{x}$ and the output $y$. There are a wide variety of spiking models based on excitability, threshold behavior, relaxation oscillations, etc. covered, for example, in \citeasnoun{Izhikevich2007}.

Physical implementations of these nonlinearities can arise from devices falling into roughly three categories: pure electronics, electro-optic physics in modulators, and active laser behavior (\figref{alt_nodes}). \Figref{alt_nodes}(a) illustrates spiking lasers, which are detailed in the next section and offer perhaps the most promise in terms of garnering the full advantage of recent theoretical results on spike processing efficiency and expressiveness. \Figref{alt_nodes}(b) is a spiking modulator. The work in \citeasnoun{VanVaerenbergh2012} might be adapted to fit this classification; however, to the authors' knowledge, an ultrafast spiking modulator remains to be experimentally demonstrated. \Figref{alt_nodes}(c) illustrates a purely electronic approach to nonlinear neural behavior. Linear E/O could be done by either a modulator or directly driven laser. This class could encompass interesting intersections with efficient analog electronic neurons in silicon~\cite{Indiveri:2011,Pickett:13}. A limitation of these approaches is the need to operate slow enough to digitize outputs into a form suitable for electronic TDM and/or AER routing.

\Figref{alt_nodes}(d) describes a laser with continuous nonlinearity, an instantiation of which was recently demonstrated in \citeasnoun{Nahmias:2016}. \Figref{alt_nodes}(e) shows a modulator with continuous nonlinearity, the first demonstration of which in a PNN and recurrent network is presented in~\cite{Tait:2017}. The pros and cons between the schemes in \figref{alt_nodes}(d) and (e) are the same ones brought up by the on-chip vs. off-chip light source debate, currently underway in the silicon photonics community. On-chip sources could provide advantageous energy scaling~\cite{Heck:14}, but they require the introduction of exotic materials to the silicon photonics process to provide optical gain. Active research in this area has the goal of making source co-integration feasible~\cite{Liang2010,Roelkens:2010}. An opposing school of thought argues that on-chip sources are still a nascent technology~\cite{Vlasov:12}. While fiber-to-chip coupling presents practical issues~\cite{Barwicz:15}, discrete laser sources are cheap and well understood. Furthermore, on-chip lasers dissipate large amounts of power~\cite{Sysak:11}, the full implications of which may complicate system design~\cite{Vlasov:12}. In either case, the conception of a PNN module, consisting of a photonic weight bank, detector, and E/O converter, as a participant in a broadcast-and-weight network could be applied to a broad array of neuron models and technological implementations.

\parornament

Both discussed all-optical and O/E/O PNN approaches depend on charge carrier dynamics, whose lifetime eventually limits the bandwidth of the summation operation. The O/E/O strategy, however, has a few advantages: it can be modularized; it uses more standard optoelectronic components; and it is more amenable to integration. Therefore here we give more attention to this strategy. Moreover, although the E/O part of the PNN can involve any kind of nonlinearity~(\figref{alt_nodes}), not necessarily spiking, we are focusing on spiking behavior because of its interesting noise-resistance and richness of representation. As such, we study here excitable semiconductor laser physics with the objective of directly producing optical spikes.

In this light, the PNN could be separated into three parts, just like the artificial neuron: weighting, addition, and neural behavior. Weighting and adding defines how nonlinear nodes can be \emph{networked} together, whereas the neural behavior dictates the \emph{activation function} shown in \figref{nonlinear_neuron_model}. In the next section, we review recent developments of semiconductor excitable lasers that emulate spiking neural behavior and, following that, we discuss a scalable WDM networking scheme.

\section*{Excitable/Spiking Lasers}
\label{sec:excitable_laser}

In the past few years, there has been a bloom of optoelectronic devices exhibiting excitable dynamics isomorphic to a physiological neuron. Excitable systems can be roughly defined by three criteria: (a) there is only one stable state at which the system can indefinitely stay at rest; (b) when excited above a certain threshold, the system undergoes a stereotypical excursion, emitting a \emph{spike}; (c) after the excursion, the system decays back to rest in the course of a \emph{refractory period} during which it is temporarily less likely to emit another spike.

\begin{figure}[!ht]
  \centering
  \includegraphics[width=0.9\linewidth]{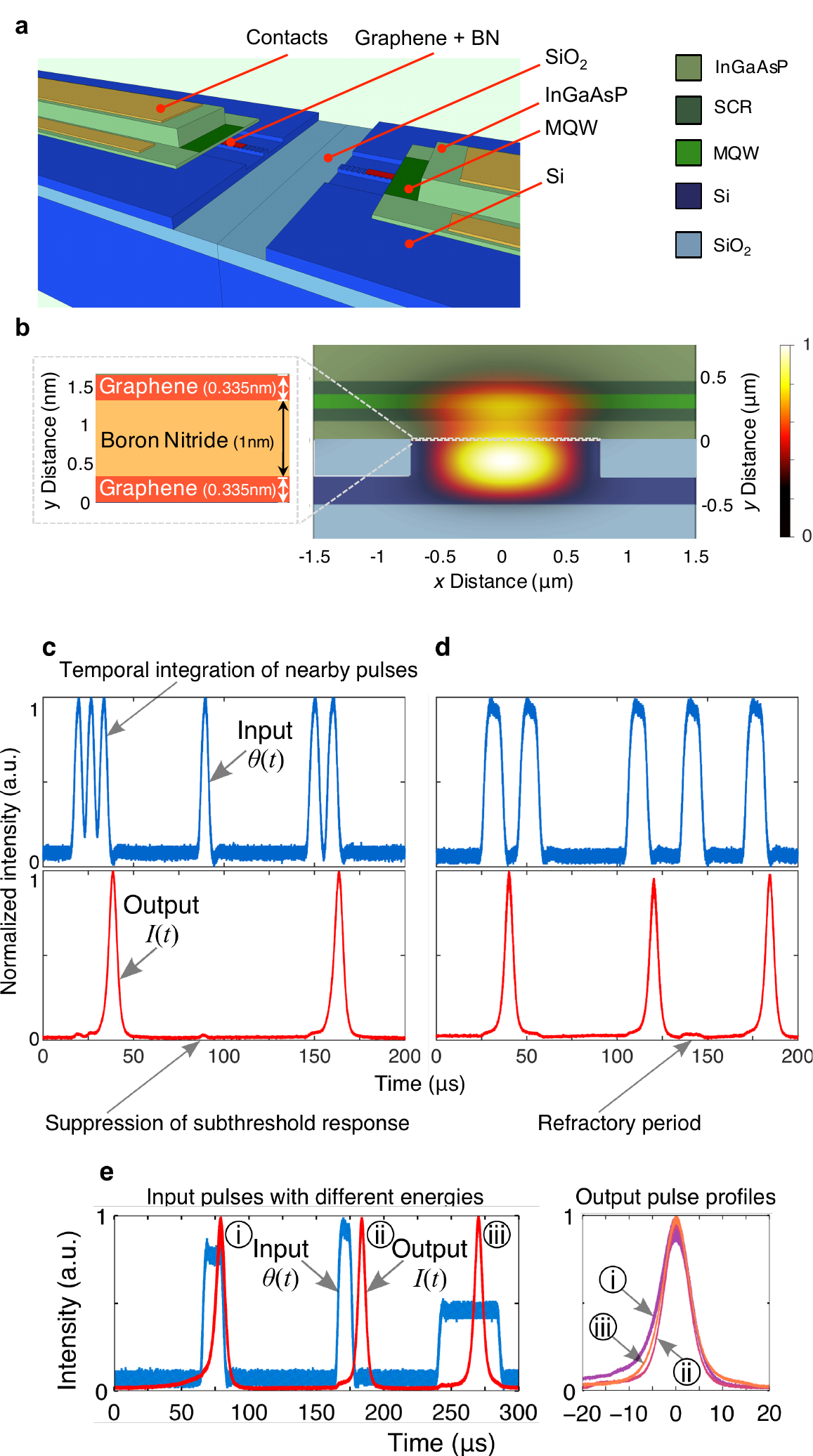}
  \caption{Excitable dynamics of the graphene excitable laser. Blue and red curves correspond to input and output pulses, respectively. (a) Cutaway architecture of a hybrid InGaAsP-graphene-silicon evanescent laser (not to scale) showing a terraced view of the center. (b) Cross-sectional profile of the excitable laser with an overlaid electric field (E-field) intensity $|\vec{E}|^2$ profile. (c--e) Excitable dynamics of the graphene \emph{fiber} laser. (c) Excitatory activity (temporal integration of nearby pulses) can push the gain above the threshold, releasing spikes. Depending on the input signal, the system can have a suppressed response due to the presence of either sub-threshold input energies (integrated power $\int |\theta(t)|^2 dt$) or (d) a refractory period during which the laser is unable to pulse (regardless of excitation strength). (e) Restorative properties: repeatable pulse shape even when inputs have different energies. Reproduced from  Shastri~\emph{et al.} \emph{Sci. Rep.} \textbf{6}, 19126 EP -- (2016) \citeasnoun{Shastri:2016aa}. Licensed under Creative Commons Attribution License (CC BY).}
  \label{fig:excitable_dynamics}
\end{figure}

\subsection*{Analogy to Leaky Integrate-and-Fire Model}
\label{subsec:lif_analogy}

Excitable behavior can be realized near the threshold of a passively Q-switched 2-section laser with saturable absorber (SA). \Figref{excitable_dynamics}(a--b) shows an example of integrated design in a hybrid photonics platform. This device comprises a III-V epitaxial structure with multiple quantum well (MQW) region (the gain region) bonded to a low-loss silicon rib waveguide that rests on a silicon-on-insulator (SOI) substrate with sandwiched layers of graphene acting as a saturable absorber region. The laser emits light along the waveguide structure into a passive silicon network. \Figref{excitable_dynamics}(c--e) shows experimental data from a fiber ring laser prototype, demonstrating key properties of excitability.

In general, the dynamics of a two-section laser composed of a gain section and a saturable-absorber (SA) can be described by the Yamada model (Eq.~\ref{eq:yamadaG}--\ref{eq:yamadaI})~\cite{Yamada:1993}. This 3-D dynamical system, in its simplest form, can be described with the following undimensionalized equations~\cite{Barbay:2011,Nahmias2013}:

\begin{IEEEeqnarray}{rCl}
  \label{eq:yamadaAll}
  \frac{\diff G(t)}{\diff t} &=& \gamma_G [ A - G(t) - G(t) I(t)] + \theta(t) \label{eq:yamadaG}\\
  \frac{\diff Q(t)}{\diff t} &=& \gamma_Q [B - Q(t) - a Q(t) I(t)\\
  \label{eq:yamadaQ}
  \frac{\diff I(t)}{\diff t} &=& \gamma_I [G(t) - Q(t) - 1] I(t) + \epsilon f(G),
  \label{eq:yamadaI}
\end{IEEEeqnarray}

where $G(t)$ models the gain, $Q(t)$ is the absorption, $I(t)$ is the laser intensity, $A$ is the bias current of the gain region, $B$ is the level of absorption, $a$ describes the differential absorption relative to the differential gain, $\gamma_G$ is the relaxation rate of the gain, $\gamma_Q$ is the relaxation rate of the absorber, $\gamma_I$ is the inverse photon lifetime, $\theta(t)$ is the time-dependent input perturbations, and $\epsilon f(G)$ is the spontaneous noise contribution to intensity; $\epsilon$ is a small coefficient.

In simple terms, if we assume electrical pumping at the gain section, the input perturbations are integrated by the gain section according to Eq.~\ref{eq:yamadaG}. An SA effectively becomes transparent as the light intensity builds up in the cavity and bleaches its carriers. It was shown in~\cite{Nahmias2013} that the near-threshold dynamics of the laser described can be approximated to Eq.~\ref{eq:lif_analogy}:

\begin{IEEEeqnarray}{rCl}
  \frac{\diff{G(t)}}{\diff{t}}&&=-\gamma_G(G(t)-A)+\theta(t); \label{eq:lif_analogy}\\
  &&\text{if $G(t)>G_{\mathrm{thresh}}$ then} \nonumber \\
  &&\text{release a pulse, and set $G(t)\rightarrow G_{\mathrm{reset}}$},\nonumber
\end{IEEEeqnarray}

where $G(t)$ models the gain, $\gamma_G$ is the gain carriers relaxation rate and $A$ the gain bias current. The input $\theta(t)$ can include spike inputs of the form $\theta(t) = \sum_i \delta_i (t - \tau_i)$ for spike firing times $\tau_i$, $G_{\mathrm{thresh}}$ is the gain threshold, and $G_{\mathrm{reset}}\around 0$ is the gain at transparency.

One can note the striking similarity to the LIF model in Eq.~\ref{eq:lif_model}: setting the variables $\gamma_G=1/R_m C_m$, $A=V_L$, $\theta(t)= I_{app}(t)/R_m C_m$, and $G(t)=V_m(t)$ shows their algebraic equivalence. Thus, the gain of the laser $G(t)$ can be thought of as a virtual \emph{membrane voltage}, the input current $A$ as a virtual \emph{equilibrium voltage} etc.

A remarkable difference can be observed between the two system though: whereas in the neural cell membrane the timescales are governed by an $R_mC_m$ constant of the order of \si{\ms}, the carrier dynamics in lasers are as fast as \si{\ns}. Although this form of excitability was found in two-section lasers, other device morphologies have also shown excitable dynamics. The advantage of constructing a clear abstraction to the LIF model is that it allows engineers to reuse the same methods developed in the computational neuroscience community for programming a neuromorphic processor.

In the next section we present recent optical devices with excitable dynamics.

\subsection*{Semiconductor Excitable Lasers}
\label{subsec:semiconductor_excitable_lasers}

\newcolumntype{b}{X}
\newcolumntype{s}{>{\hsize=.5\hsize}X}
\newcolumntype{l}{>{\hsize=.3\hsize}X}

\begin{table*}[!ht]
  \caption{Characteristics of recent excitable laser devices. Note that this table does not have a one-to-one correspondence to \figref{alt_nodes}, because some of them are not E/O devices. However, we observe that devices A, D, and F belong to the category \figref{alt_nodes}(a), and device E, which resembles more closely to the category \figref{alt_nodes}(c).}
  \begin{tabularx}{\textwidth}{bslss}%{bslss}
    \hline\hline
    \textbf{Device}   & \textbf{Injection Scheme}   & \textbf{Pump}   & \textbf{Excitable Dynamics}   & \textbf{Refs.}\\
    \hline
    A. Two-section gain and SA                          & electrical          & electrical  & stimulated emission   & \cite{Spuhler:99,Dubbeldam1999,Dubbeldam:1999a,Larotonda:2002,Elsass:2010,Barbay:2011,Nahmias2013,Shastri:OQE:2014,Selmi:2014,Nahmias:2015,Selmi:15,Shastri:15,Shastri:2016aa}\\
    B. Semiconductor ring laser                         & coherent optical    & electrical  & optical interference  & \cite{Coomans2010,Gelens2010a,Coomans2011,VanVaerenbergh2012,Coomans2013}\\
    C. Microdisk laser                                  & coherent optical    & electrical  & optical interference  & \cite{VanVaerenbergh2013,Alexander2013}\\
    D. 2D Photonic crystal nanocavity\footnotemark[1]   & electrical          & electrical  & thermal   & \cite{Yacomotti:2006,Yacomotti:2006b,Brunstein:2012}\\
    E. Resonant tunneling diode photodetector and laser diode\footnotemark[2]   & electrical or incoherent optical  & electrical  & electrical tunneling  & \cite{Romeira:2013,Romeira:2016aa}\\
    F. Injection-locked semiconductor laser with delayed feedback               & electrical  & electrical  & optical interference  & \cite{Wieczorek1999,Wieczorek2002,Barland2003,Wieczorek2005,Marino2005,Goulding2007,Kelleher2010,Kelleher2011,Turconi:2013,Garbin2014,Garbin2015}\\
    G. Semiconductor lasers with optical feedback       & incoherent optical  & electrical  & stimulated emission   & \cite{Giudici:1997,Yacomotti:1999,Giacomelli2000,Heil:2001,Wunsche:2001,Aragoneses:2014,Sorrentino:15}\\
    H. Polarization switching VCSELs                    & coherent optical    & optical     & optical interference  & \cite{Hurtado2010a,Hurtado2012,Hurtado:APL:2015}\\
    \hline\hline
  \end{tabularx}
  \footnotetext[1]{Technically this device is not an excitable laser, but an excitable cavity connected to a waveguide.}
  \footnotetext[2]{The authors call it \emph{excitable optoelectronic device}, because the excitability mechanism lies entirely in an electronic circuit, rather than the laser itself.}
  \label{table:excitable_devices}
\end{table*}

Optical excitability in semiconductor devices are being widely studied, both theoretically and experimentally. These devices include multisection lasers, ring lasers, photonic crystal nanocavities, tunneling diode attached to laser diodes, and semiconductor lasers with feedback, summarized in Table~\ref{table:excitable_devices}. We group them under the terminology \emph{excitable lasers} for convenience, but exceptions are described in the caption of the table.

Generally speaking, these lasers use III-V quantum wells or quantum dots for efficient light generation. However, they fall into one of three injection categories (illustrated in \figref{excitable_lasers}) and possess very diverse excitability mechanisms. It is difficult to group the rich dynamics of different lasers---which often requires a system of several coupled ordinary differential equations to represent it---using classification keywords.
We focus on two fundamental characteristics: the the way each laser can be modulated (injection scheme column), and on the physical effect that directly shapes the optical pulse (excitable dynamics column).

The injection scheme of the laser will determine whether it is compatible to all-optical PNNs or O/E/O PNNs. Some of them (B, C, H) operate free of electrical injection, meaning that bits of information remain elegantly encoded in optical carriers. However, as we have point out, avoiding the E/O conversion is much more difficult when you are trying to build an weight-and-sum device compatible with WDM, which is an essential building block for scalable photonic neural networks.

The excitable dynamics determines important properties such as energy efficiency, switching speed and bandwidth of the nonlinear node. The ``optical interference'' mechanism typically means that there are two competing modes with a certain phase relationship that can undergo a $2\pi$ topological excursion and generating an optical pulse in amplitude at the output port. This mechanism is notably different than the others in which it does not require exchange of energy between charge carriers populations and the cavity field. As a result, systems based on this effect are not limited by carrier lifetimes, yet are vulnerable to phase noise accumulation. Other mechanisms include photon absorption, stimulated emission, thermo-optic effect, and electron tunneling. There, the electronic dynamics of the device governs the population of charge carriers available for stimulated emission, thereby dominating the time scale of the generated pulses. Models of these mechanisms and how they elicit excitability are comprehensively detailed in \citeasnoun{Prucnal:16}, but a quantitative comparison between performance metrics of lasers in Table~\ref{table:excitable_devices} is still called for. Qualitatively, however, excitable lasers can simultaneously borrow key properties of electronic transistors, such as thresholding and cascadability).

In addition to individual laser excitability, there have been several demonstrations of simple processing circuits. Temporal pattern recognition~\cite{Shastri:2016aa} and stable recurrent memory~\cite{Romeira:2016aa,Garbin2014,Shastri:2016aa} are essential toy circuits that demonstrate basic aspects of network-compatibility.

\begin{figure*}[!ht]
  \centering
  \includegraphics[width=0.75\linewidth]{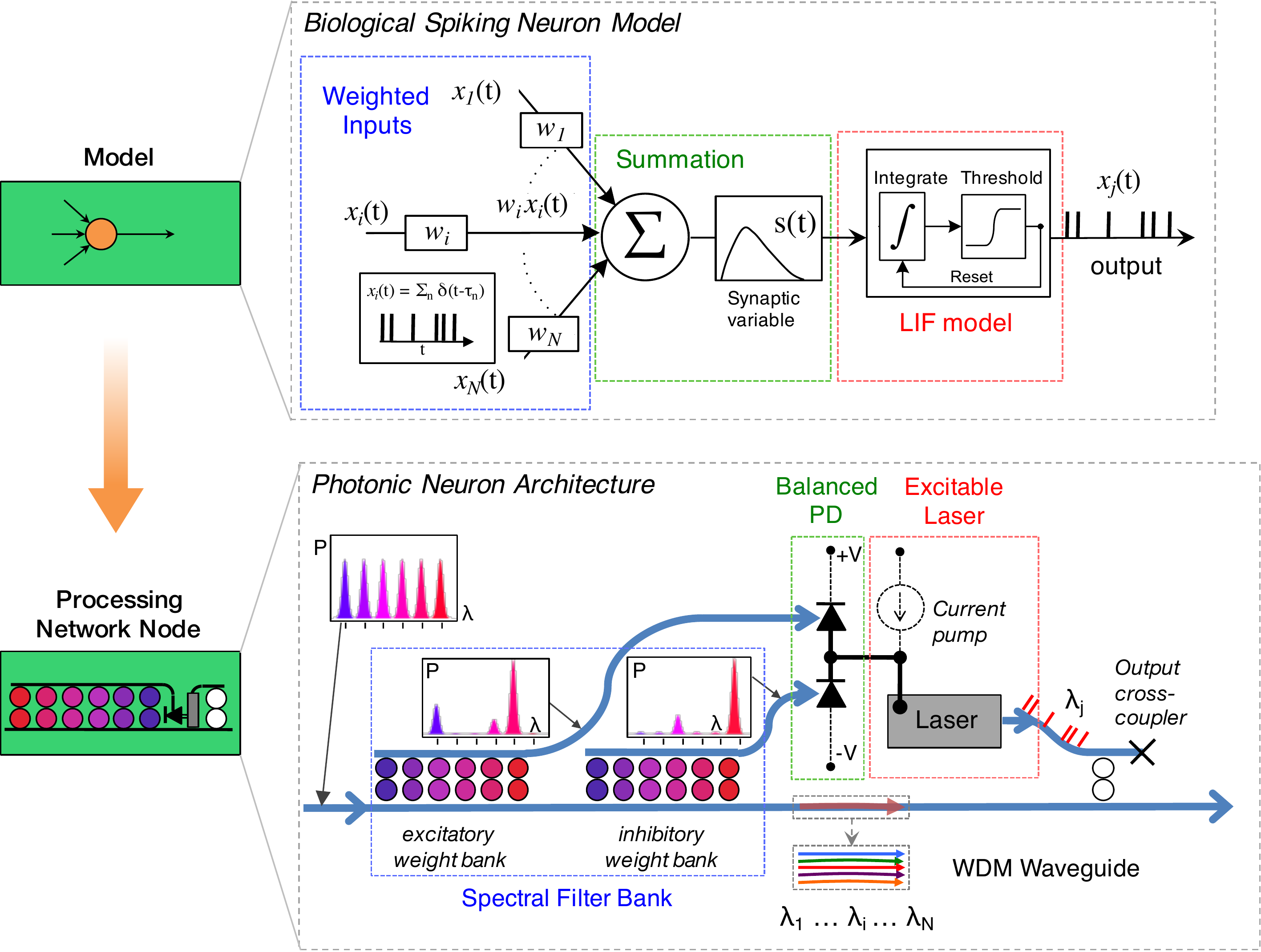}
  \caption{Isomorphism between photonic neuron module (i.e., a PNN) to a biological spiking neuron model. Top: depiction of a single unit neural network model. Inputs $x_i$ are weighted and summed. The result $\sum_i w_ix_i$ experiences a nonlinear function. Bottom: Possible physical implementation of a PNN. A WDM signal is incident on a bank of filters (excitatory and inhibitory)---which are created using a series of microring filters---apply a series of weights. The resulting signal is incident on a balanced photodetector which applies a summation operation and drives an excitable laser with a current signal. The resulting laser outputs at a specified wavelength which is subsequently coupled back into the broadcast interconnect (see \figref{broadcast_weight}). Adapted with permission from Tait~\emph{et al.} \emph{J. Lightwave Technol.} \textbf{32}, 4029--4041 (2014) \citeasnoun{Prucnal:16}. Copyright 2014 Optical Society of America.}
  \label{fig:pnn}
\end{figure*}

\section*{Photonic Neural Network Architecture}
\label{sec:photonic_neural_networks}

\subsection*{Isomorphism to Biological Spiking Neuron}
\label{subsec:isomorphism}

Neurons only have computational capabilities if they are in a network. Therefore, an excitable laser (or spiking laser) can only be viewed as a neuron candidate if it is contained in a PNN (\figref{pnn}). The configurable analog connection strengths between neurons, called weights, are as important to the task of network processing as the dynamical behavior of individual elements. Earlier, we have discussed several proposed excitable lasers exhibiting neural behavior and cascadability between these lasers. In this section, we discuss the challenges involving the creation of a network of neurons using photonic hardware, in particular, the creation of a weighted addition scheme for every PNN. Tait~\etal.~\cite{Tait:JLT:2014} proposed an integrated photonic neural networking scheme called \emph{broadcast-and-weight} that uses WDM to support a large number of reconfigurable analog connections using silicon photonic device technology.

\begin{figure}[!ht]
  \centering
  \includegraphics[width=0.9\linewidth]{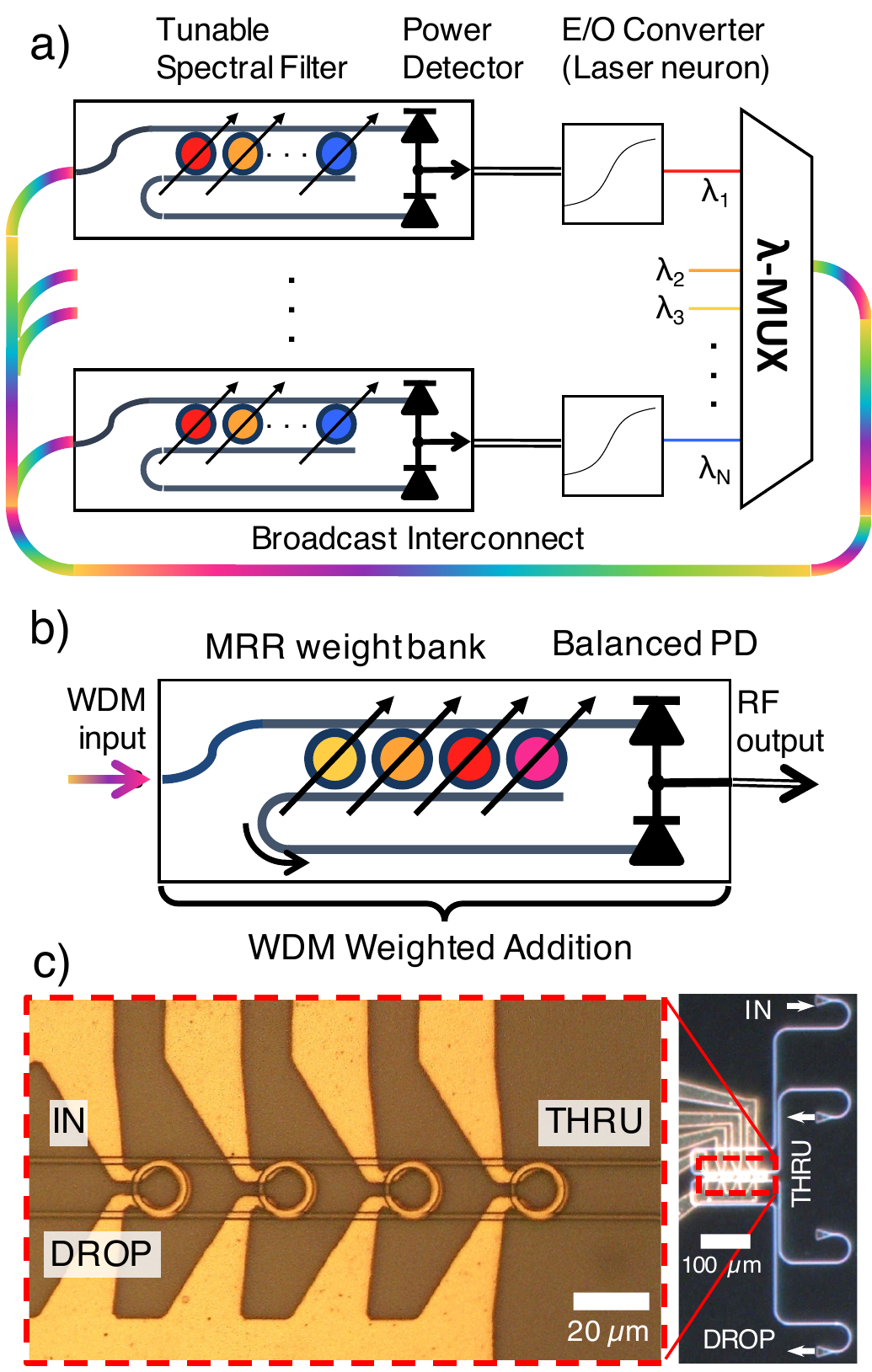}
  \caption{(a) Broadcast-and-weight network. An array of source lasers outputs distinct wavelengths (represented by solid color). These channels are wavelength multiplexed (WDM) in a single waveguide (multicolor). Independent weighting functions are realized by tunable spectral filters at the input of each unit. Demultiplexing does not occur in the network. Instead, the total optical power of each spectrally weighted signal is detected, yielding the sum of the input channels. The electronic signal is transduced to an optical signal after nonlinear transformation. (b) Tunable spectral filter constructed using microring resonator (MRR) weight bank. Tuning MRRs between on- and off-resonance switches a continuous fraction of optical power between drop and through ports. A balanced photodetector (PD) yields the sum and difference of weighted signals. (c) \emph{left}: Optical micrograph of a silicon MRR weight bank, showing a bank of four thermally-tuned MRRs. \emph{right:} Wide area micrograph, showing fiber-to-chip grating couplers. Reproduced with permission from Tait~\emph{et al.} \emph{Opt. Express} \textbf{24}, 8895--8906 (2016) \citeasnoun{Tait:16multi}. Copyright 2016 Optical Society of America.}
  \label{fig:broadcast_weight}
\end{figure}

A spiking and/or analog photonic network consists of three aspects: a protocol, a node that abides by that protocol (the PNN), and a network medium that supports multiple connections between these nodes. This section will begin with broadcast-and-weight as a WDM protocol in which many signals can coexist in a single waveguide and all nodes have access to all the signals. Configurable analog connections are supported by a novel device called a microring resonator (MRR) weight bank (\figref{broadcast_weight}). We will summarize experimental investigations of MRR weight banks.

\subsection*{Broadcast-and-Weight Protocol}
\label{subsec:broadcast_and_weight}

WDM channelization of the spectrum is one way to efficiently use the full capacity of a waveguide, which can have usable transmission windows up to \SI{60}{nm} (\SI{7.5}{\THz} bandwidth)~\cite{Preston:11}. In fiber communication networks, a WDM protocol called broadcast-and-\emph{select} has been used for decades to create many potential connections between communication nodes~\cite{Ramaswami:1993}. In broadcast-and-select, the active connection is selected, not by altering the intervening medium, but rather by tuning a filter at the receiver to drop the desired wavelength. Broadcast-and-\emph{weight} is similar, but differs by directing multiple inputs simultaneously into each detector (\figref{broadcast_weight}(b)) and with a continuous range of effective drop strengths between --1 and +1, corresponding to an analog weighting function.

The ability to control each connection, each weight, independently is a crucial aspect of neural network models. Weighting in a broadcast-and-weight network is accomplished by a tunable spectral filter bank at each node, an operation analogous to a neural weight. The local state of the filters defines the interconnectivity pattern of the network.

A great variety of possible weight profiles allows a group of functionally similar units to instantiate a tremendous variety of neural networks. A reconfigurable filter can be implemented by a MRR---in simple words, a waveguide bent back on itself to create an interference condition. The MRR resonance wavelength can be tuned thermally (as in \figref{broadcast_weight}(c)) or electronically on timescales much slower than signal bandwidth. Practical, accurate, and scalable MRR control techniques are a critical step towards large scale analog processing networks based on MRR weight banks. %We present them in Section~\ref{subsec:weight-bank-control}. Analysis of scaling and design for MRR weight banks is then given in Section~\ref{subsec:weight-bank-analysis}.

\subsection*{Controlling Photonic Weight Banks}
\label{subsec:weight_bank_control}

Sensitivity to fabrication variations, thermal fluctuations, and thermal cross-talk has made MRR control an important topic for WDM demultiplexers~\cite{Klein:05}, high-order filters~\cite{Mak:15}, modulators~\cite{Cox:14}, and delay lines~\cite{Cardenas:10}. Commonly, the goal of MRR control is to track a particular point in the resonance relative to the signal carrier wavelength, such as its center or maximum slope point. On the other hand, an MRR weight must be biased at arbitrary points in the filter roll-off region in order to multiply an optical signal by a continuous range of weight values. Feedback control approaches are well-suited to MRR demultiplexer and modulator control~\cite{DeRose:10, Jayatilleka:15opex}, but these approaches rely on having a reference signal with consistent average power. In analog networks, signal activity can depend strongly on the weight values, so these signals cannot be used as references to estimate weight values. These reasons dictate a feedforward control approach for MRR weight banks.

\subsubsection*{Single Channel Control Accuracy and Precision}
\label{subsubsec:single-channel_precision}
How accurate can a weight be? The resolution required for effective weighting is a topic of debate within the neuromorphic electronics community, with IBM's TrueNorth selecting four digital bits plus one sign bit~\cite{Akopyan:15}. In \citeasnoun{Tait:15cont}, continuous weight control of an MRR weight bank channel was shown using an interpolation-based calibration approach. The goal of the calibration is to have a model of applied current/voltage vs. effective weight command. The calibration can be performed once per MRR and its parameters can be stored in memory. Once calibration is complete, the controller can navigate the MRR transfer function to apply the correct weight value for a given command. However, errors in the calibration, environmental fluctuations or imprecise actuators causes the weight command to be inaccurate. It is necessary to quantify that accuracy.

Analog weight control accuracy can be characterized in terms of the ratio of weight range (normalized to 1.0) to worst-case weight inaccuracy over a sweep and stated in terms of bits or a dynamic range. The initial demonstration reported in \citeasnoun{Tait:15cont} indicates a dynamic range of the weight controller of 9.2dB, in other words, an equivalent digital resolution of 3.1 bits.

\subsubsection*{Multi-Channel Control Accuracy and Precision}
\label{subsubsec:multi_channel_precision}
Another crucial feature of an MRR weight bank is simultaneous control of all channels. When sources of cross-talk between one weight and another are considered, it is impossible to interpolate the transfer function of each channel independently. Simply extending the single-channel interpolation-based approach of measuring a set of weights over the full range would require a number of calibration measurements that scales exponentially with the channel count, since the dimension of the range grows with channel count. Simultaneous control in the presence of cross-talk therefore motivates model-based calibration approaches.

Model-based, as opposed to interpolation-based, calibration involves parameterized models for cross-talk-inducing effects. The predominant sources of cross-talk are thermal leakage between nearby integrated heaters and, in a lab setup, inter-channel cross-gain saturation in fiber amplifiers, although optical amplifiers are not a concern for fully integrated systems that do not have fiber-to-chip coupling losses. Thermal cross-talk occurs when heat generated at a particular heater affects the temperature of neighboring devices (see \figref{broadcast_weight}(c)). In principle, the neighboring channel could counter this effect by slightly reducing the amount of heat its heater generates. A calibration model for thermal effects provides two basic functions: forward modeling (given a vector of applied currents, what will the vector of resultant temperatures be?) and reverse modeling (given a desired vector of temperatures, what currents should be applied?). Models such as this must be calibrated to physical devices by fitting parameters to measurements. Calibrating a parameterized model requires at least as many measurements as free parameters. \citeasnoun{Tait:16multi} describes a method for fitting parameters with $O(N)$ spectral and oscilloscope measurements, where $N$ is the number of MRRs. As an example, whereas an interpolation-only approach with 20 points resolution would require $20^4 = 160,000$ calibration measurements, the presented calibration routine takes roughly $4 \times \left[10 (\mbox{heater}) + 20 (\mbox{filter}) + 4 (\mbox{amplifier})\right] = 136$ total calibration measurements. Initial demonstrations achieved simultaneous 4-channel MRR weight control with an accuracy of 3.8 bits and precision of 4.0 bits (plus 1.0 sign bit) on each channel. While optimal weight resolution is still a topic of discussion in the neuromorphic electronics community~\cite{Hasler2013}, several state-of-the-art architectures with dedicated weight hardware have settled on 4-bit resolution~\cite{Friedmann:13, Akopyan:15}.

\subsection*{Scalability with Photonic Weight Banks}
\label{subsec:weight_bank_analysis}

Engineering analysis and design relies on quantifiable descriptions of performance called metrics. The natural questions of ``how many channels are possible'' and, subsequently, ``how many more or fewer channels are garnered by a different design'' are typically resolved by studying tradeoffs. Increasing the channel count performance metric will eventually degrade some other aspect of performance until the minimum specification is violated.

Studying tradeoffs between these metrics are important for better designing the network and understanding its limitations. Just as was the case with control methodologies, it was found that quantitative analysis for MRR weight banks must follow an approach significantly different from those developed for MRR demultiplexers and modulators~\cite{Tait:16anal}.

In conventional analyses of MRR devices for multiplexing, demultiplexing, and modulating WDM signals, the tradeoff that limits channel spacing is inter-channel cross-talk~\cite{Preston:11,Jayatilleka:15}. However, unlike MRR demultiplexers where each channel is coupled to a distinct waveguide output~\cite{Klein:05}, MRR weight banks have only two outputs with some portion of every channel coupled to each. All channels are meant to be sent to both detectors in some proportion, so the notion of cross-talk between signals breaks down (Fig.~\ref{fig:broadcast_weight}(b)). Instead, for dense channel spacing, different filter peaks viewed from the common drop port begin to merge together. This has the effect of reducing the weight bank's ability to weight neighboring signals independently. As detailed in \citeasnoun{Tait:16anal}, Tait~\etal (1) quantify this effect as a power penalty by including a notion of tuning \emph{range} with the cross-weight penalty metric, and (2) perform a channel density analysis by deriving the scalability of weight banks that use microresonators of a particular finesse.

\begin{figure*}[!ht]
  \centering
  \includegraphics[width=0.8\linewidth]{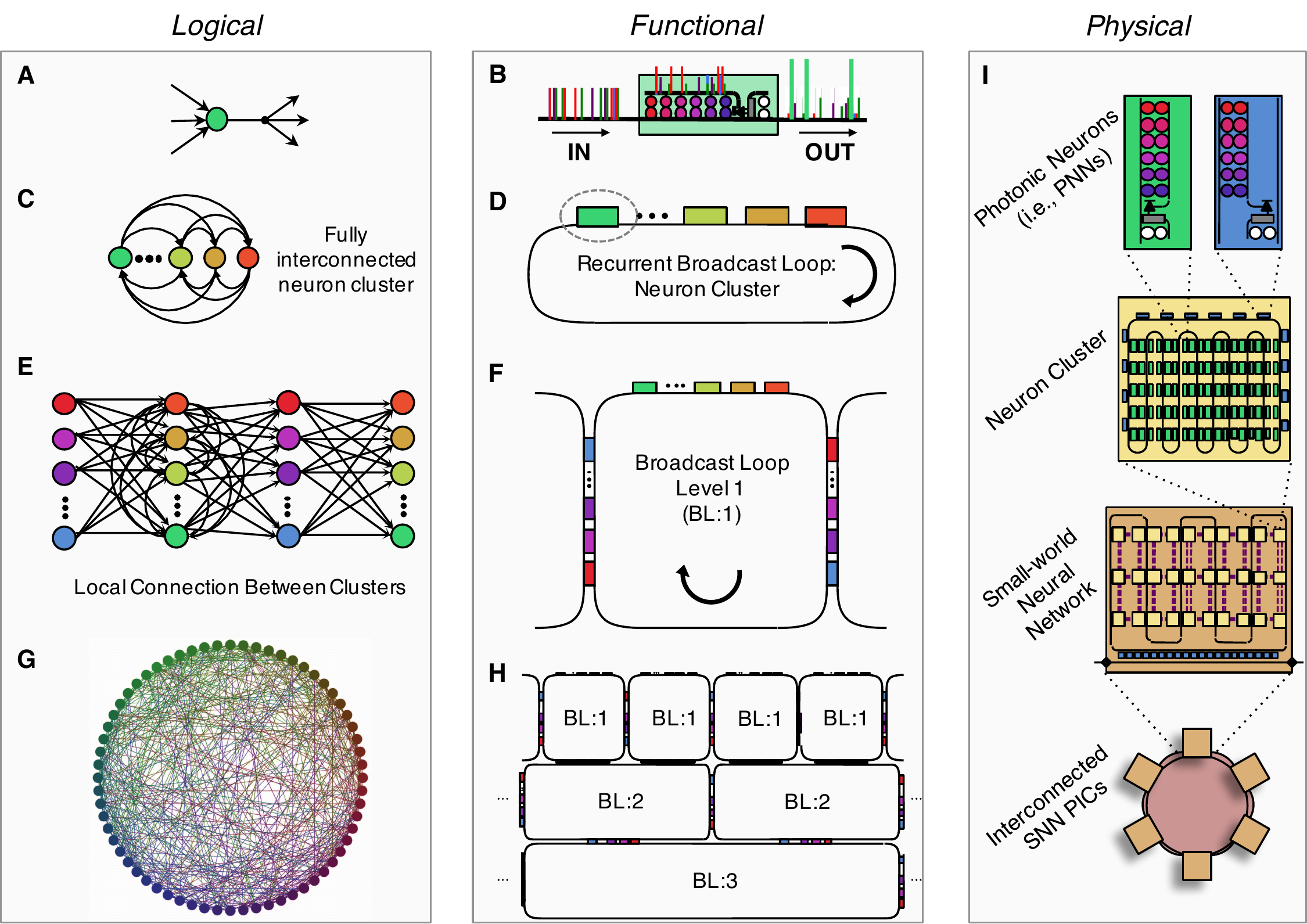}
  \caption{Spectrum reuse strategy. Panels are organized into rows at different scales (core, cluster, chip, and multichip) and into columns at three different views (logical, functional, and physical). (a,b) Depicts the model and photonic implementation of a neuron as a processing network node (PNN) as detailed in \figref{pnn}. (c,d) Fully interconnected network by attaching PNNs to a broadcast loop (BL) waveguide. (e,f) A slightly modified PNN can transfer information from one BL to another. (g,h) Hierarchical organization of the waveguide broadcast architecture showing a scalable modular structure. (i) Using this scheme, neuron count in one chip is only limited by footprint, but photonic integrated circuits (PICs) can be further interconnected in an optical fiber network.}
  \label{fig:bl_hierarchy}
\end{figure*}

In summary, WDM channel spacing, $\delta \omega$, can be used to determine the maximum channel count given a resonator finesse. While finesse can vary significantly with the resonator type, normalized spacing is a property of the circuit (i.e. multiplexer vs. modulators vs. weight bank). Making an assumption that a 3dB cross-weight penalty is allowed, we find that the minimum channel spacing falls between 3.41 and 4.61 linewidths depending on bus length. High finesse silicon MRRs, such as shown in \citeasnoun{Xu:2008} (finesse = 368) and~\cite{Biberman:12} (finesse = 540), could support 108 and 148 channels, respectively. Other types of resonators in silicon, such as elliptical microdisks~\cite{Xiong:11} (finesse = 440) and traveling-wave microresonators~\cite{Soltani:10} (finesse = 1,140) could reach up to 129 and 334 channels, respectively.

\parornament

MRR weight banks are an important component of neuromorphic photonics -- regardless of PNN implementation –– because they control the configuration of analog network linking photonic neurons together. In \citeasnoun{Tait:16multi}, it was concluded that ADC resolution, sensitivized by biasing conditions, limited the attainable weight accuracy. Controller accuracy is expected to improve by reducing the mismatch between tuning range of interest and driver range. \citeasnoun{Tait:16anal} arrived at a scaling limit of 148 channels for a MRR weight bank, which is not impressive in the context of neural networks. However, the number of neurons could be extended beyond this limit using spectrum reuse strategies (\figref{bl_hierarchy}) proposed in~\cite{Tait:JLT:2014}, by tailoring interference within MRR weight banks as discussed in~\cite{Tait:16anal}, or by packing more dimensions of multiplexing within silicon waveguides, such as mode-division multiplexing. As the modeling requirements for controlling MRR weight banks become more computationally intensive, a feedback control technique would be transformative for both precision and modeling demands. Despite the special requirements of photonic weight bank devices making them different from communication-related MRR devices, future research could enable schemes for feedback control.

\newcolumntype{b}{X}
\newcolumntype{s}{>{\hsize=.5\hsize}X}
\begin{table*}
  \caption{Comparison Between Different Neuromorphic Processors}
  \label{tab:comparison_neuromorphic}
  \begin{tabularx}{\textwidth}{bsssss}
    \hline\hline
    \textbf{Chip}   & \textbf{MAC rate/processor\footnotemark[3]}   & \textbf{Energy/MAC (\si{\pico\joule})}\footnotemark[4]  & \textbf{Processor fan-in}   & \textbf{Area/MAC (\si{\um^2})}\footnotemark[5]  & \textbf{Synapse precision (bit)}\\
    \hline
    Photonic Hybrid III-V/Si\footnotemark[1]                  & \SI{20}{\GHz}   & \num{0.26}     & 108         & 205   & 5.1\\
    Sub-$\lambda$ Photonics \footnotemark[2] (future trend)   & \SI{200}{\GHz}  & \num{0.0007}  & $\sim$200   & 20    & 8\\
    HICANN~\cite{Schemmel2010}                                & \SI{22.4}{\MHz} & \num{198.4}   & 224         & 780   & 4\\
    TrueNorth~\cite{Merolla08082014}                          & \SI{2.5}{\kHz}  & \num{.27}     & 256         & 4.9   & 5\\
    Neurogrid~\cite{Benjamin:2014}                            & \SI{40.1}{\kHz} & \num{119}     & 4096        & 7.1   & 13\\
    SpiNNaker\footnotemark[6]~\cite{Furber:2014}              & \SI{3.2}{\kHz}  & \num{6e5}     & 320         & 217   & 16\\
    \hline\hline
  \end{tabularx}
  \footnotetext[1]{III-V/Si Hybrid stands for estimated metrics of a spiking neural network in a photonic integrated circuit in a III-V/Si Hybrid platform.}
  \footnotetext[2]{Sub-$\lambda$ stands for estimated metrics for a platform using optimized sub-wavelength structures, such as photonic crystals.}
  \footnotetext[3]{A MAC event occurs each time a spike is integrated by the neuron. Neuron fan-in refers to the number of possible connections to a single neuron.}
  \footnotetext[4]{The energy per MAC for HICANN, TrueNorth, Neurogrid, SpiNNaker were estimated by dividing wall-plug power to number of neurons and to operational MAC rate per processor.}
  \footnotetext[5]{The area per MAC was estimated by dividing the chip/board size to the number of MAC units (neuron count times fan-in). All numbers therefore include overheads in terms of footprint and area.}
  \footnotetext[6]{Neurons, synapses and spikes are digitally encoded in event headers that travel around co-integrated processor cores. So all numbers here are based on a typical application example.}
\end{table*}

\section*{Neuromorphic platforms comparison}
\label{sec:platforms_comparison}

As stated earlier, the neuromorphic computing community has been making vigorous efforts toward large-scale spiking neuromorphic hardware, e.g., Heidelberg HICANN chip via the FACETS/BrainScaleS projects~\cite{Schemmel2010} IBM TrueNorth via the DARPA SyNAPSE program~\cite{Merolla08082014}, Stanford’s Neurogrid~\cite{Benjamin:2014} and SpiNNaker~\cite{Furber:2014} (Fig.~\ref{fig:spike_platforms}). Many researchers are concentrating their efforts towards the long-term technical potential and functional capability of the hardware compared to standard digital computers. One of the main drivers for the community is computational power efficiency~\cite{Hasler2013}: digital CPUs are reaching a power efficiency wall with the current von Neumann architecture, but hardware neural networks, in which memory and instructions are simplified and colocated, offer to overcome this barrier.

These projects also aimed at simulating large-scale spiking neural networks, with the goal of simulating subcircuits of the human cortex, at a biological timescale ($<\SI{1}{\kHz}$). The HICANN chip, exceptionally, is designed to be accelerated at about 10 thousand times respective to biological time scales, and feature analog synapses and realistic neural spiking behaviors~\cite{Schemmel2010}. It pays the price of huge power consumption: \SI{800}{\watt} for a wafer-scale system containing 180,000 neurons~\cite{Benjamin:2014}. In contrast, TrueNorth aimed for large-scale, efficient networks optimized for biologically plausible tasks, such as machine vision, but with a simplified neural model~\cite{Merolla08082014}. Indeed, it contained a total of 1M neurons and 256M synapses per chip, consuming only \SI{63}{\milli\watt} of power~\cite{Merolla08082014}. The Neurogrid board also aimed for scalability and efficiency, consuming \SI{5}{\watt} also with 1M neurons and about 4 billion synapses, but it kept greater biological fidelity to the mammalian cortex. The SpiNNaker computer is designed to simulate scalably large and versatile networks using arrays of chips with 18 ARM968 processing cores each: unlike the other three technologies, the number of synapses per neuron, or even the number of neurons, is not fixed and can be dynamically reprogrammed~\cite{Furber:2014}. The demonstrated system has 48 chips interconnected in a PCB, but it can be scaled up to 1200 interconnected PCBs, totalling a \SI{72}{\kilo\watt} peak power consumption.

We have recently produced a quantitative comparison between the aforementioned electronic and photonic neuromorphic hardware architectures~\cite{PrucnalBook}. In order to compare these processors with one another, we reintroduce the multiply-accumulate (MAC) operation that typically bottlenecks complex computations.

%The gap in computational efficiency between the biological neuron and the current general-purpose digital circuits is very large. An important measure for computational efficiency is the MAC/Joule, introduced in Chapter~\ref{chapterNeuroEngineering}. While the efficiency of physiological neurons is estimated to be more than $10^{18}$ MACs/Joule, the energy-efficiency wall of current processors is $10^{12}$ MACs/Joule, or six orders of magnitude lower~\cite{Hasler2013}.

%Weighted addition is critical for neural network implementations, and---since the number of operations scales quadratically with the number of nodes in all-to-all connected networks---it represents the most costly hardware scalability bottleneck~\cite{Hasler2013}. Thus for analysis, we can deconstruct this operation as a parallelized set of multiply and accumulates (MACs) and use it as a reference unit of computation.

%\subsection*{Multiply-Accumulate Operations}
%\label{subsec:macs}

The MAC operation takes the following form: $a \leftarrow a+(w\times x)$. It includes both a product (i.e. $x$ is multiplied by the `weight' $w$) and an addition (the result is accumulated to variable $a$). In neural network models, inputs are combined via a weighted sum of the form $\sum w_i x_i$. The result is then input into some nonlinear scalar function $f\{x\}$ which can range from a simple sigmoid function to a complex nonlinear dynamical system with hysteresis, depending on the complexity of the neuron being modeled. The weighted sum can be broken down into a series of MAC operations of the form $a_i = a_{i-1}+w_i x_i$ for $i = 1\ldots M$. Each neuron requires $M$ parallel MAC operations per time step $\Delta t$ (in a given bit period $\tau$, determined by the signal bandwidth capacity), or one operation per synapse, where $M$ refers to the number of inputs for a given node. Thus, a hardware neural network can be characterized with $M \times N$ MAC operations per time step $\Delta t$ (i.e., quadratic scaling of MAC operations), where $N$ is the number of neurons in the network.

The nonlinear function $f\{x\}$ also takes up computational resources, but since this operation scales with $N$ rather than $M\times N$, it does not represent the most costly operation. Therefore, as the size of the network $N$ grows large, MACs---i.e. `synaptic computations'---become the most burdensome hardware bottlenecks in neural networks~\cite{Hasler2013}.

For consistency, we compare architectures that have similar functionality: we limit ourselves to fully reconfigurable systems of spiking neural networks. %The analysis includes electronic neuromorphic architectures introduced in earlier.
For the photonically-enhanced system, we studied an optoelectronic neural network with PNNs instantiated within the hybrid Silicon/III-V platform~\cite{Nahmias:2015,FerreiradeLima:16sumtop}. We also consider a future photonic crystal instantiation, based on fundamental physical considerations. Calculated metrics are based on realistic device parameters, derived from the literature. We also refer the reader to a more detailed discussion of spiking electronic neuromorphic hardware in~\cite{Liu2015} and an overview of current spiking and non-spiking hardware in~\cite{Nawrocki2016}.

Results are summarized in Table~\ref{tab:comparison_neuromorphic}. The most striking figure is the number of operations per second, which exceeds electronic platforms by 3 orders of magnitude, compared to the analog/digital accelerated HICANN and 6 orders of magnitude compared to the others which are purely digital implementations. This stems from both the high bandwidths and low latencies possible with photonic signals. The optoelectronic approach is able to achieve such energy efficiency at high speeds because power is mainly consumed statically by the lasers, while the passive filters have low leakage current. This contrasts with CMOS digital switches, whose power consumption increases dynamically with clock speed.
Processor fan-in is similar in both platforms, despite very differing technologies. The area per MAC is more stringent in a photonically enhanced system, since photonic elements cannot be shrunk beyond the diffraction limit of light. This is because each data channel requires a weighting filter in the PNN, such as an MRR pair, which adds a footprint penalty. However, this is compensated by the fact that a single waveguide can carry many wideband channels simultaneously, unlike electronic wires. Nonetheless, even though photonically enhanced systems cannot compete with the miniaturization of future nanoelectronics, the estimated footprint of such a system is currently on par with some of the electronics systems presented here.

\section*{Future Directions}
\label{sec:future_direction}
%(a discussion including potential impacts on the development of certain areas of science)

\begin{figure}[!ht]
  \centering
  \includegraphics[width=0.85\linewidth]{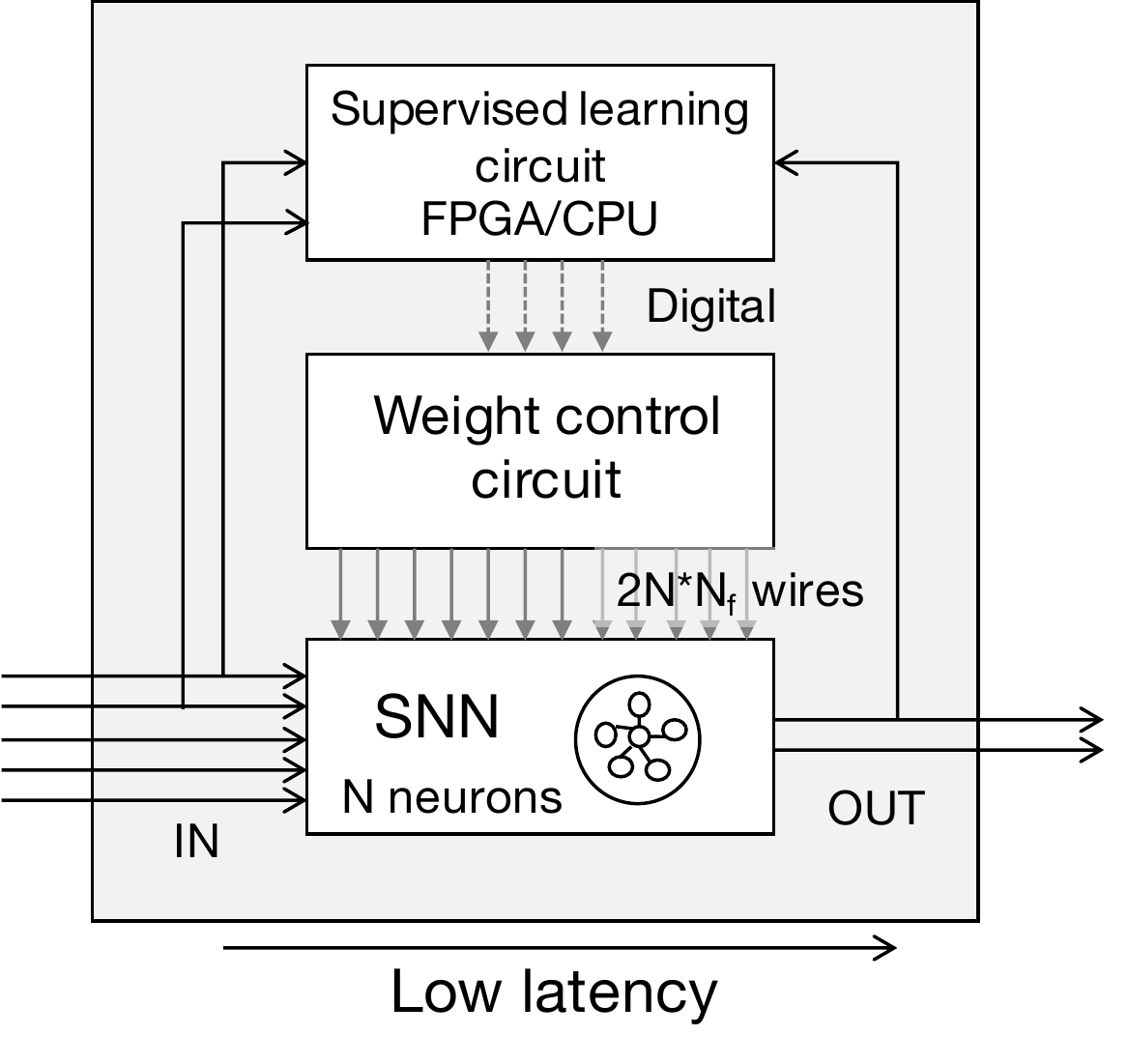}
  \caption{Diagram description of a fully packaged neuromorphic processor. While two layers of electronics provide reconfigurability, the photonic spiking neural network permits low-latency functionality. $N_f$: Fan-in of each neuron. Reproduced from Ferreira de Lima~\emph{et al.} \emph{Nanophotonics} \textbf{6}, 577--599 (2017) \citeasnoun{FerreiradeLima:2017}. Licensed under Creative Commons Attribution-NonCommercial-NoDerivatives License. (CC BY-NC-ND).}
  \label{fig:neuromorphic_processor}
\end{figure}

After half a century of continuous investment and commercial success, digital CMOS electronics dominates the industry of general-purpose computing.
However, with growing demand for connectivity, there is an urgent need for ultrafast co-processors that could relieve the stress in digital processing circuits. Here, we have presented elements of a reconfigurable photonic hardware that can emulate spiking neural networks operating a billion times faster than the brain. As we identify proper metrics for a neuromorphic photonic processor, research efforts are incipiently transitioning from individual devices to systems design. We are witnessing a fast maturation of standardized photonic foundries in several platforms. Chrostowski \& Hochberg say~\cite{Chrostowski:15} we are entering a nascent era of fabless photonics, where users can create computer-assisted chip designs and have it fabricated by these foundries using quality controlled, repeatable processes. It is reasonable to expect that neuromorphic photonic co-processors (\figref{neuromorphic_processor}) can be fabricated and packaged using fabless services for near-term research and long-term volume production.

Applications for neuromorphic photonic processors can be grouped into two categories: 1. a frontend stage for RF systems and data centers; and 2. ultrafast processing for specialized fast applications~\cite{PrucnalBook}. The first category utilizes the low-latency, parallelism, and energy efficiency properties of photonics to alleviate the throughput of RF systems, e.g. by executing dimensionality reduction tasks such as principal component analysis or blind-source separation. The second category takes advantage of the raw speed (bandwidth and latency) of the photonic processor to execute iterative algorithms mapped to recurrent neural networks.

Neuromorphic photonic processors join a class of photonic hardware accelerators designed to assist in acquisition, feature extraction, and storage of wideband waveforms~\cite{Jalali2015}. These accelerators manipulates the spectrotemporal of a wideband signal, task difficult to accomplish in analog electronics over broad bandwidth and with low loss.

\subsection*{Real-Time Radio Frequency Processing}
\label{subsec:rf_processing}

High volume data applications, including streaming video and cloud services, will continue to push the telecommunications industry to build better, high bandwidth systems. Data traffic on some mobile networks alone has increased by over 6000\%~\cite{Cisco:2014}. This has motivated the exploration of more efficient usage of spectral resources~\cite{Akyildiz2006}. Although RF integrated circuits (RFICs) have been researched for applications such as duplex processing~\cite{lee2003design,razavi2000design} or control of beamforming antennas~\cite{yu2011integrated,Razavi:2009}, the requirement for impedance-matched transmission lines greatly increases device and interconnect footprint, limiting the overall complexity of each chip. Photonics provides a solution to these fundamental limitations: optical waveguides can support large bandwidths (\around100s of THz) with high information density and low crosstalk between multiplexed channels. By using techniques such as WDM, a large number of multi-GHz channels can exist within the same optical waveguide. As a result, the number of virtual channels can greatly exceed the number of physical waveguides, allowing for the formation of complex processing circuits without a significant hardware overhead.

%Cognitive radios (CR)~\cite{Haykin:2005} have evolved as a facilitating technology by allowing users to intelligently sense the RF environment and adaptively exploit available spectral holes~\cite{Akyildiz:2006}. However, typical RF devices such as software radios, are limited in bandwidth (\aroundMHz range). Microwave electronics, in contrast, can operate on higher bandwidths, but are strongly frequency dependent, and exhibit poor wideband performance and reconfigurability. Hardware neural network approaches are promising for radio cognition. However, because large fan-in is a necessity for neural networks, electronic versions are severely limited in speed with signals in the kHz or slower. In the GHz regime, electronic fan-in is strongly bounded by impedance mismatching, RC limitations, and cross talk between closely spaced wires.

\begin{figure*}[!ht]
  \centering
  \includegraphics[width=0.75\linewidth]{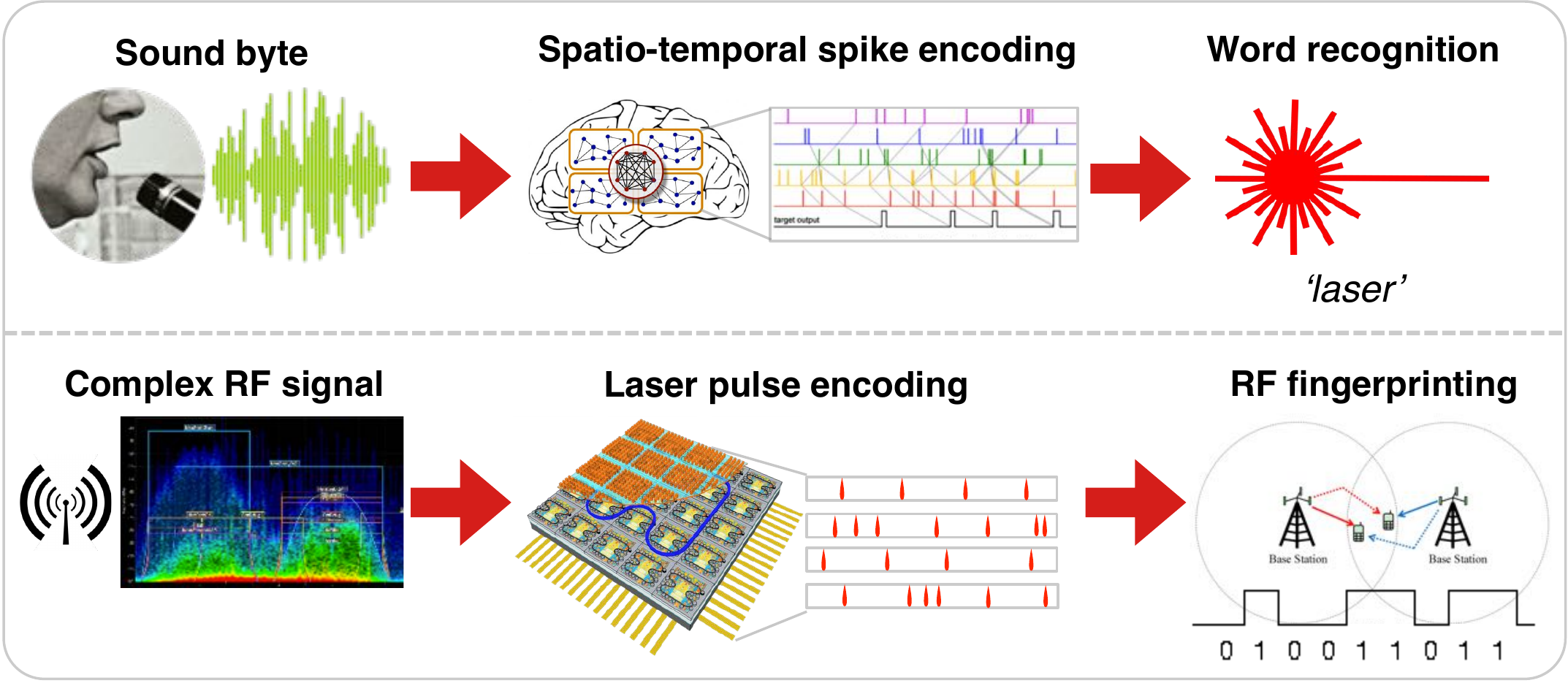}
  \caption[A vision for a brain-inspired laser neural network for enhanced RF communication]{A vision for a brain-inspired laser neural network for enhanced RF communication. Top: spoken words are transformed into spatio-temporal `spike' (event) patterns by biological neurons. A pattern recognition neuron is sensitive to a specific spike fingerprint and releases its own spike only if it occurs, shown in `target output'. Bottom: in analogy with audio waveforms, a much faster system (\around GHz) based on excitable lasers could operate directly on RF waveforms. Applying operations at the front-end of RF transceivers could offload complex signal processing operations to a photonic chip, and address bandwidth and latency limitations of current FPGA and DSP solutions. The spike coded pattern is reproduced from Tapson~\emph{et al.} Front. Neurosci. \textbf{7} (2013) \citeasnoun{Tapson2013}. Licensed under Creative Commons Attribution License (CC BY).}
  \label{fig:cr_application}
\end{figure*}

%Due to their large bandwidth and small footprint, a semiconductor-based realization may provide the prospect of scalability that is required to investigate ultrafast processors.
%The idea is to build a fundamental platform---an on-chip optical architecture to support parallel communication among a network of excitable lasers (\figref{cr_application}) and demonstrate a basic implementation of pattern recognition of an incoming analog RF signal.
After some initial front-end processing (i.e., heterodyning and amplification), most radio transceiver systems are processed by either digital signal processors (DSP) or field programmable gate arrays (FGPAs) for more complex signal operations. However, the speeds of these processors (i.e., \around\SI{500}{MHz}) limit the overall throughput of RF carrier signals, which can easily be in the \around GHz. Clever sampling and parallelization can help alleviate this bottleneck, but at the cost of much higher latency and a significant resource/energy overhead. Specialized RF application-specific integrated circuits (ASICs) are another option, but are expensive, require significant development time and have limited reconfigurability. Future imagined multiple-in multiple-out (MIMO) systems---which in the case of massive MIMO, can be on the order of \around100s of input and output channels~\cite{Larsson:2014,Gesbert:2003}---are especially susceptible to this bottleneck, and may require a radically new solution.

Adding a photonic processing chip to the front of a radio transceiver would allow very complex operations to be performed in real-time, which can significantly offload electronic post-processing and provide a technology to make faster, more relevant RF decisions on-the-fly. Massive MIMO systems based on beamforming in phased array antennas require a processor that can distinguish and operate on hundreds of high bandwidth signals simultaneously, a feat that is currently speed limited by current electronic processors~\cite{Larsson:2014,hansen2009phased}. A photonic neural network model is a perfect fit for addressing this kind of technological challenge:  efficient MIMO beamforming relies on MAC operations that are already applied in neural network models via \emph{weighted addition}. In addition, classification algorithms can be built efficiently using the neural network approach, allowing for RF fingerprinting and signal identification.

As spread spectrum, adaptive RF transceivers become more widespread in future telecommunication systems, the scalability of the photonic approach could provide significant processing advantages. Its high bandwidth, low latency, and high throughput would be especially useful in an ultra-wideband (UWB) radio system, in which it could sample from many frequencies and directions simultaneously to scan for spectral opportunities and make a decision quickly and efficiently.  Pairing this technology with an FPGA or electronic ASIC controller would enable the implementation of adaptive optimization and learning algorithms in real time for ultrafast cognitive radio applications.
%These spike processing circuits can be implemented in conventional device fabrication processes made for silicon: manufacturing process with billions of dollars of commercial investment. The incorporation of spike processing circuits into photonic hardware---through the exploitation of both the high bandwidth of optical signals and the dynamics of excitable lasers---could potentially bring the computational intelligence of neural networks into the hitherto unexplored GHz regime.

\subsection*{Nonlinear Programming}
\label{subsec:nonlinear_programming}

Another way of taking advantage of raw speed is via an \emph{iterative} approach. Iterative algorithms find successfully better approximations to a problem of interest, and often require many time steps to reach a desired solution.
%Therefore, the latency of the underlying hardware can play a critical role in determining the convergence rate of these algorithms.
Since one of the most salient advantages of a photonic approach is its low time-of-flight (\around\si{ps}) between communicating processors, the convergence rates can be significantly improved by implementing them on a photonic platform.  A large class of problems that can be solved iteratively includes \emph{linear and nonlinear programming problems.} These methods seek to minimize some objective function $E(\vec{x})$ of real variables in $\vec{x}$ subject to a series of constraints represented by equalities or inequalities i.e. $g(\vec{x}) \le 0$ and $h(\vec{x}) = 0$.
%Whether or not the problem is \emph{linear} is based on the linearlity of the objective function.
Applications in telecommunications, aerospace, and financial industries can be described in this basic framework, including optimal portfolio trading strategies, control of machinery/actuators, and allocation of resources and jobs in online servers. Using a photonic approach, 100-variable problems could converge in less than \around$\SI{100}{ns}$, which could be useful in the control of very fast dynamical systems (i.e., actuators) or in the creation of low-latency optimization routines in data-intensive environments.

Mathematical optimization problems can be grouped into \emph{linear} and \emph{nonlinear} optimization problems. Nonlinear optimization problems are often difficult to solve, and sometimes involve exotic techniques such as genetic algorithms or particle swarm optimization. Nonlinear optimization problems, however, are nonetheless \emph{quadratic} to second order around the local vicinity of the optimum. Therefore, Quadratic Programming (QP)---which finds the minima/maxima quadratic functions of variables subject to linear constraints~\cite{Lendaris:1999}---becomes an effective first pass at such problems, and can be applied to a wide array of applications.
%Quadratic Programming (QP) is a subset of mathematical optimization which aims to find minima of quadratic functions of variables subject to linear constraints~\cite{Lendaris:1999}. QP is a generalized mathematical optimization tool which has applications in a wide variety of fields.
For example, many machine learning problems, such as support vector machine (SVM) training and least squares regression, can be reformulated in terms of a QP problem. In addition, computational problems such as model predictive control (MPC), an optimal nonlinear control algorithm, or compressive sampling, a method for sampling at rates below the Nyquist without loss of information via the characterization of sparsity in incoming data, are examples of QP problems. Together, these applications represent some of the most effective yet generalized tools for acquiring and processing information and using the results to control systems. QP is an NP hard problem in the number of variables, which means that conventional digital computers must either be limited to solving quadratic programs of very few variables, or to applications where computation time is noncritical. This is reflected in industrial applications of QP solvers. MPC is used in the chemical industry to control chemical processing plants, where reaction timescales can be made very long, and in the finance industry to control long term portfolio optimization. The application of MPC to faster systems, therefore, relies on new ways of finding faster solutions to QP problems~\cite{Jerez2011}. In machine learning, many algorithms (such as SVM) require offline training because of the computational complexity of QP, but would be much more effective if they could be trained online.

%It has been shown that quadratic programs can be mapped onto recurrent neural networks that converge to an attractor state corresponding to the solution of QPs~\cite{Xia2001}.

\begin{figure}[!ht]
  \centering
  \includegraphics[width=1.0\linewidth]{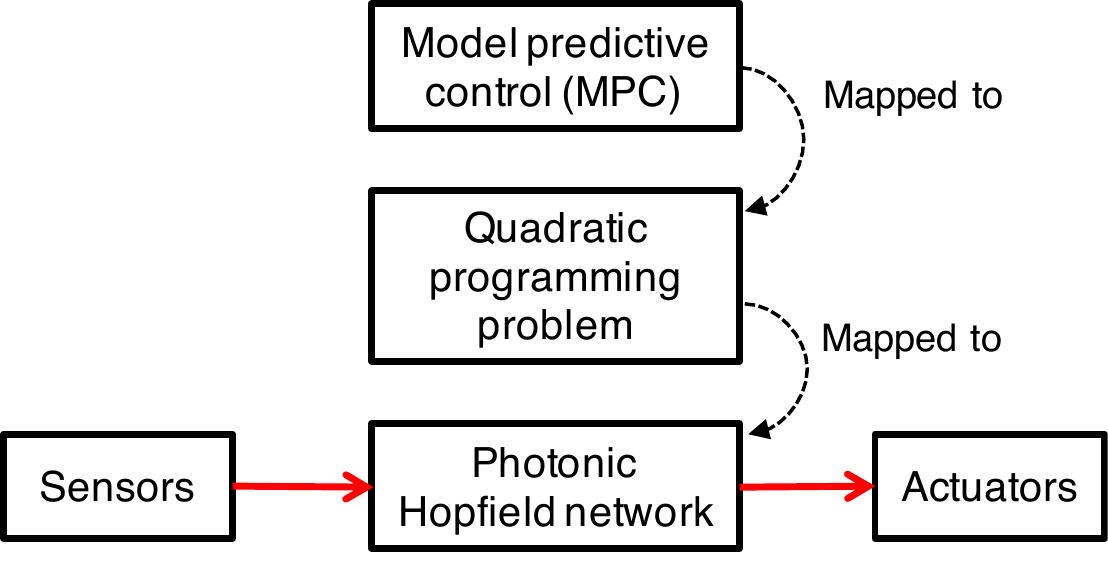}
  \caption{Employing a photonic neural network for a model predictive control problem by solving a quadratic programming problem with a Hopfield network.}
  \label{fig:mpc}
\end{figure}

While Hopfield showed that Hopfield Networks are able to solve quadratic optimization problems quickly over 25 years ago~\cite{Tank:1986}, Hopfield quadratic optimizers are uncommon today. This is largely due to the high connectivity between neurons that neural networks require ($n^2$ connections for $n$ neurons). In an electronic circuit, as the number of connections increases, the bandwidth at which the system can operate without being subject to crosstalk between connections and other issues decreases~\cite{Tait:JLT:2014}. This creates an undesirable tradeoff between neuron speed and neural network size. Photonic neural networks have several advantages over their electrical counterparts. Most importantly, the connectivity concerns prevalent in electronic neurons are significantly ameliorated by using light as a communication medium~\cite{Tait:JLT:2014}. WDM allows for hundreds of high bandwidth signals to flow through a single optical waveguide. Moreover, the analog computational bandwidth of a photonic neuron (as designed in Tait~\etal \citeasnoun{Tait:15cont}) lies in the picosecond to femtosecond time scale~\cite{Nahmias:2015}. For a Hopfield Quadratic Optimizer, this means that a photonic implementation (\figref{mpc}) can simultaneously have large dimensionality and a fast convergence time to the minimum. These processors represent some of the most effective, yet generalized tools for acquiring and processing information and controlling highly-mobile systems, such as a hypersonic aircraft~\cite{Keviczky2006}.

\section*{Summary and Conclusion}
\label{subsec:summary_conclusion}

Photonics has revolutionized information communication, while electronics, in parallel, has dominated information processing. Recently, there has been a determined exploration of the unifying boundaries between photonics and electronics on the same substrate, driven in part by Moore's Law approaching its long-anticipated end. For example, the computational efficiency for digital processing has leveled-off around \SI{100}{\pico\joule} per MAC. As a result, there has been a widening gap between today's computational efficiency and the next generation needs, such as big data applications which require advanced pattern matching and real-time analysis. This, in turn, has led to expeditious advances in: (1) emerging devices that are called ``beyond-CMOS'' or ``More-than-Moore'', (2) novel processing or unconventional computing architectures called ``beyond-von Neumann'', that are brain-inspired, i.e., neuromorphic, and (3) CMOS-compatible photonic interconnect technologies. Collectively, these research endeavors have given rise to the field of neuromorphic photonics (\figref{landscape}). Emerging photonic hardware platforms have the potential to vastly exceed the capabilities of electronics by combining ultrafast operation, moderate complexity, and full programmability, extending the bounds of computing for applications such as navigation control on hypersonic aircrafts and real-time analysis of the RF spectrum.

In this article, we discussed the current progress and requirements of such a platform. In a photonic spike processor, information is encoded as events in the temporal and spatial domains of spikes (or optical pulses). This hybrid coding scheme is digital in amplitude but analog in time, and benefits from the bandwidth efficiency of analog processing and the robustness to noise of digital communication. Optical pulses are received, processed, and generated by certain class of semiconductor devices that exhibit excitability---a nonlinear dynamical mechanism underlying all-or-none responses to small perturbations. Optoelectronic devices operating in the excitable regime are dynamically analogous to a biophysical neuron, but roughly eight orders of magnitude faster. We dubbed these devices as ``photonic neurons'' or ``laser neurons.'' The field is now reaching a critical juncture where there is a shift from studying single photonic neurons to studying an interconnected networks of such devices. A recently proposed on-chip networking architecture called broadcast-and-weight could support massively parallel (\emph{all-to-all}) interconnection between excitable devices using wavelength division multiplexing.

A hybrid III-V/Si photonics platform is a candidate for an integrated hardware platform. III-V compound semiconductor technology, such as indium phosphide (InP) and gallium arsenide (GaAs), is at the forefront of providing \emph{active} elements like lasers, amplifiers and detectors. Silicon, in parallel, brings compatibility with CMOS fabrication processes and low-loss \emph{passive} components like waveguides and resonators. Scalable and fully reconfigurable networks of excitable lasers can be implemented in the silicon photonic layer of modern hybrid integration platforms, in which spiking lasers in a bonded InP layer are densely interconnected through a silicon layer. Such a photonic spike processor will potentially be able to support several thousand interconnected devices. It is predicted that such a chip would have a computational efficiency of \SI{260}{\femto\joule} per MAC, which surpasses the energy efficiency wall by two orders of magnitude while operating at high speeds (i.e. signal bandwidths \SI{10}{\GHz}). The emerging field of photonic spike processors has received tremendous interest and continues to develop as photonic integrated circuits increase in performance and scale. As novel applications requiring real-time, ultrafast processing---such as the exploitation of the RF spectrum---become more demanding, we expect that these systems will find use in a variety of high performance, time-critical environments.

Moving forward, we envision a tremendous interest in designing, building and understanding photonic networks of excitable elements for ultrafast information processing, guided by the latest computational models of the brain. Successful implementation of a small-scale photonic spike processor could, in principle, provide the fundamental technology to build and study larger scale brain-inspired networks based on laser excitability. Neuromorphic photonics is poised to usher in exciting new fields of inquiry and impactful enterprises of application.

%\begin{acknowledgments}
%\end{acknowledgments}

\bibliography{neuroprinciples}% Produces the bibliography via BibTeX.

\end{document}